\documentclass[12pt]{article}
\usepackage{epsfig, amssymb}
\usepackage{amsmath}
\usepackage[normalem]{ulem}
\useunder{\uline}{\ul}{}
\usepackage[title]{appendix}
\usepackage{graphicx,epsfig}
\usepackage{booktabs, colortbl, array}
\usepackage{color} 
\usepackage{cite}
\usepackage{verbatim}
\usepackage{caption}
\begin{document}
%\begin{titlepage}
\thispagestyle{empty}
\begin{flushright}
\end{flushright}

\bigskip
\begin{center}
\noindent{\Large \textbf
{Holographic entanglement entropy probe on spontaneous symmetry breaking with vector order%asymptotically AdS and anisotropic spacetime
}}\\ 
%Supersonic Flows and their Gravity Duals
\vspace{2cm}  
\noindent{Chanyong Park${}^{a}$\footnote{e-mail:cyong21@gist.ac.kr}, Gitae Kim${}^{b}$\footnote{e-mail:kimgitae728@gmail.com}, Ji-seong Chae${}^{b}$\footnote{e-mail:mp3dp98@hanyang.ac.kr } and 
Jae-Hyuk Oh${}^{b}$\footnote{corresponding author, e-mail:jaehyukoh@hanyang.ac.kr}}

\vspace{1cm}
  {\it
Department of Physics and Photon Science, Gwangju Institute of Science and Technology,
Gwangju 61005, Korea${}^{a}$\\
Department of Physics, Hanyang University, Seoul 04763, Korea${}^{b}$\\
 }
\end{center}

\vspace{0.3cm}
\begin{abstract}
\noindent
%We explore a certain 4-dimensional dual field theory near the critical point  by employing holographic entanglement entropy, where the field theory system shows second order phase transition by spontaneous symmetry breaking with vector order parameter.
We study holographic entanglement entropy in 5-dimensional charged black brane geometry obtained from Einstein-SU(2)Yang-Mills theory defined in asymptotically AdS space. This gravity system undergoes second order phase transition near its critical point, where %affected by 
a spatial component of the Yang-Mills fields appears, which is normalizable mode of the solution. This is known as phase transition between isotropic and anisotropic phases, where in anisotropic phase, SO(3)-isometry(spatial rotation) in bulk geometry is broken down to SO(2) by emergence of the spatial component of Yang-Mills fields, which corresponds to a vector order in dual field theory. 
%In dual field theory, this corresponds to spontaneous symmetry breaking with vector order parameter.
%The dual gravity model that we study is Einstein-SU(2)Yang-Mills theory defined in asymptotically AdS space. 
We get analytic solutions of holographic entanglement entropies by utilizing the solution of bulk spacetime geometry given in arXiv:1109.4592, where we consider subsystems defined on AdS boundary of which shapes are wide and thin slabs and a cylinder.
% on the AdS boundary. 
%For the slabs, we consider the slab being parallel to the vector order and perpendicular to the vector order.with same shape but stting in different orientations and .
It turns out that the entanglement entropies near the critical point shows scaling behavior such that for both of the slabs and cylinder, $\Delta_\varepsilon S\sim\left(1-\frac{T}{T_c}\right)^\beta$ and %it turns out that 
the critical exponent $\beta=1$, where $\Delta_\varepsilon S\equiv S^{iso}-S^{aniso}$, and $S^{iso}$ denotes the entanglement entropy in isotropic phase whereas $S^{aniso}$ denotes that  in anisotropic phase. We suggest a quantity $O_{12}\equiv S_1-S_2$ as a new order parameter near the critical point, where $S_1$ is entanglement entropy when the slab is perpendicular to the direction of the vector order whereas $S_2$ is that when the slab is parallel to the vector order. $O_{12}=0$ in isotropic phase but in anisotropic phase, the order parameter becomes non-zero showing the same scaling behavior. Finally, we show that even near the critical point, the first law of entanglement entropy is held. Especially, we find that the entanglement temperature for the cylinder is $\mathcal T_{cy}=\frac{c_{ent}}{a}$, where $c_{ent}=0.163004\pm0.000001$ and $a$ is the radius of the cylinder.

\end{abstract}

%\end{titlepage} 
\newpage
\tableofcontents

\section{Introduction}
\label{Introduction}
%Anisotropic holographic superfluids\cite{Basu:2011tt,Oh:2012zu,Park:2016wch}. \\
%Entanglement entropy\cite{Solodukhin:2008dh,Hung:2011xb}.
AdS/CFT correspondence has shed light on strongly coupled field theories by employing their holographic dual gravity theories\cite{Aharony:1999ti, Maldacena:1997re,Oh:2011wpl}. Especially, fluid-gravity duality\cite{Kovtun:2004de,Benincasa:2006fu,Iqbal:2008by} and AdS-condensed matter theory(AdS/CMT)\cite{Gubser:2008px,Hartnoll:2008kx,Gubser:2008zu,Basu:2008bh} are widely studied in many literatures to explore low energy(long wavelength) limits of conformal field theories, which become conformal fluids, condensed matter systems and so on. Especially, in fluid-gravity duality, holographic computation of the ratio of shear viscosity, $\eta$ to entropy density, $\mathcal S$ is the most remarkable example and it is known to be universal, which is given by $\frac{\eta}{\mathcal S}=\frac{1}{4\pi}$\cite{Kovtun:2004de,Iqbal:2008by,Hartnoll:2009sz,Herzog:2009xv,Horowitz:2010gk}.

An interesting issue related to fluid-gravity duality and AdS/CMT is thermodynamic phase transition where the system shows symmetry breaking because of emergence of an order parameter, a condensation. In condensed matter theory, electron-electron bound states, so called Cooper pairs, are present near its critical point which breaks $U(1)$ gauge symmetry of electrons.

{A noticeable construction of gravity model for holographic condensed matter theory is based on a theory with complex scalar field defined in the background of (asymptotically AdS) charged black brane.\cite{Hartnoll:2008kx,Gubser:2008px}. When the black brane temperature becomes below a certain critical temperature, $T=T_c$, the complex scalar field becomes {unstable and} condensed. The charged black brane geometry presents scalar hair due to this condensation. The condensation corresponds to $U(1)$-symmetry breaking due to emergence of a scalar order in the dual field theory system, and which implies super-conductor/normal-conductor phase transition.
}

It is also interesting to consider the emergence of vector(p-wave) or tensor(d-wave) order near the critical point.
%It is rather interesting to study phase transition, where the order parameter is scalar but it is vector(p-wave) or tensor(d-wave).
%Super/normal conductor phase transition is 
%Super ffluids/normal fluids phase transition is 
An interesting holographic model to explore p-wave super-fluid/normal-fluid phase transition is Einstein-SU(2)Yang-Mills theory in asymptotically AdS$_5$ spacetime\cite{Herzog:2009ci,Basu:2011tt,Oh:2012zu,Park:2016wch}.
A precise solution in the theory is a 5-dimensional charged black brane solution. In the solution, temporal direction of the Yang-Mills fields is turned on, which is proportional to $\tau_3=\frac{\sigma_3}{2}$, where $\tau_3$ is the third generator in SU(2) gauge group, $\sigma_i$ are the Pauli-matrices and $i=1,2,3$. 

In this dual gravity model, for a certain chemical potential $\mu_c$, an interesting mode of solution appears. This mode is the spatial component of Yang-Mills fields being proportional to $\tau_1=\frac{\sigma_1}{2}$ and it is a normalizable mode of the solution. In fact, the black brane geometry enjoys SO(3)-global rotational symmetry mixing 3-dimensional spatial coordinates, $\{x_1,x_2,x_3\}$. However, once the spatial mode of solution appears, the SO(3) rotational symmetry is broken down to SO(2), where the direction of the spatial mode is chosen to be in $x_1$-axis. In holographic dictionary, normalizable mode of solution in dual gravity corresponds to a state in the boundary field theory. Therefore, the symmetry is broken due to a state in the dual field theory and so it is spontaneous symmetry breaking(SSB).

One of the previous works to explore this SSB near the critical point is a study on the ratio of shear viscosity to entropy density in fluid-gravity duality\cite{Erdmenger:2010xm,Basu:2011tt}. The shear viscosity defined in $x_2-x_3$ plane, $\eta_{23}$ retains its universal value of the ratio since SO(2)-rotational symmetry still exists whereas that defined in the plane which contains $x_1$ coordinate(e.g. $\eta_{12}$) will give the deviation from the universal value. 

{
To study this holographic fluid system, one can apply either numerical or analytic methods. In \cite{Erdmenger:2010xm}, the authors consider numerical method and find out that when $\alpha=\frac{\kappa_5}{g}$ is less than a certain critical value, $\alpha_{crit}$, the boundary fluid system shows second order phase transition whereas if $\alpha$ is greater than $\alpha_{crit}$, it presents first order phase transition, where $\kappa_5$ is 5-dimensional gravity constant and $g$ is the gauge coupling of the Yang-Mills fields. They also figure out that the deviation of the ratio from the universal value shows scaling behaviors near the critical point, such as
\begin{equation}
1-4\pi\frac{\eta_{12}}{\mathcal S}\sim \left(1-\frac{T}{T_c}\right)^\beta
\end{equation}
and its critical exponent $\beta$ is 
\begin{equation}
\beta=1.00\pm 0.03,
\end{equation}
where $T$ is the charged black brane temperature, and $T_c$ is critical temperature.
}

To determine the critical exponent more precisely, the authors in \cite{Basu:2011tt} employ analytic approach in large gauge coupling limit, $\alpha\ll 1$, together with the magnitude of spatial component of Yang-Mills fields, $\varepsilon \ll 1$ is small. The deviation is given by
\begin{equation}
1-4\pi\frac{\eta_{12}}{\mathcal S}=\frac{1305\pi T_c}{544}\left(1-\frac{T}{T_c}\right)^\beta.
\end{equation}
%where $\alpha^2=\frac{\kappa_5^2}{g^2}$, $\kappa_5$ is 5-dimensional gravity constant, $g$ is SU(2) Yang-Mills coupling, $T$ is the charged black brane temperature, and $T_c$ is critical temperature. 
It turns out that the critical exponent is determined to be $\beta=1$ in this analytic approach. 
%%%%%%%%%%%%%%Entnglement entropy review%%%%%%%%%%%%%%%%

{The analytic approach given in\cite{Basu:2011tt} to obtain such a deviation of the ratio of shear viscosity to entropy density is crucially  based on their methodology of double expansion with parameters, {$\alpha$ and $\varepsilon$ in the bulk}. The authors in \cite{Basu:2011tt} assume that the magnitude of the vector order, $\varepsilon$ is small and so its subleading corrections are suppressed by $\varepsilon$. To get analytic form of the leading backreactions from the energy momentum tensor of the Yang-Mills fields, they also assume that $\alpha$ is small otherwise it is very unlikely to get the analytic solutions of the back reactions. 
This approach allows them to obtain the analytic form of the backreactions perturbatively.
%the subleading corrections of the backreactions to the background spacetime to become smaller than the leading backreaction. 
The leading corrections to the background geometry is $O(\varepsilon^2 \alpha^2)$ and the subleading corrections are $O(\varepsilon^{1+i}\alpha^{2j})$, where $i,j$ are positive integers and $i>1$ or $j>1$. We note that since this analytic approach is based on small $\alpha$ expansion, it is manifest that the thermodynamic phase transition observed in that analytic approach is second order one.
}

On the other hand, there is another interesting direction to explore the field theory system, (holographic) entanglement entropy. Entanglement entropy, which describes quantum correlation between a macroscopic 
subsystem and its complement, is one of the important quantities specifying quantum nature of a system. 
%{{%Understanding the quantum nature of strongly interacting systems is one of the important issues in nuclear and condensed matter physics. Despite the importance, it is still hard to figure out nonpertubative quantum states of strongly interacting systems because of the absence of well-established nonperturbative methods. In this situation, recently, there was an interesting proposal called the AdS/CFT correspondence or holography. It claimed that a $(d+1)$-dimensional classical gravity has a one-to-one map to a $d$-dimensional quantum field theory (QFT). This AdS/CFT correspondence provides us a new way to understand nonperturbative features of QFT. 
Based on the AdS/CFT correspondence, Ryu and Takayanagi propose that the entanglement of a 
quantum field theory can be evaluated by calculating the area of a minimal surface extending to the dual geometry 
\cite{Ryu:2006bv,Ryu:2006ef} and it is further developed in \cite{Solodukhin:2008dh,Hung:2011xb}. This provides more precise understanding between gravity theory in asymptotically AdS space and the dual field theory defined on its conformal boundary \cite{Casini:2011kv}.

%By using the holographic description of the entanglement entropy, it was shown that the $c$-function of a two-dimensional QFT monotonically decreases along the RG flow. This work was further generalized to $a$- and $F$-theorem for higher dimensional QFT. The $c$-theorem represents how the degrees of freedom of a system reduces as the RG scale decreases. 
%In the present work, we investigate how anisotropy modifies the quantum entanglement of a system. Since anisotropy breaks the rotational symmetry, quantum entanglement generally leads to different values depending on directions of the entangling region. We further study how anisotropic quantum entanglement changes under the RG flow. }}

%%%%%%%%%%%%%%%%%%%%%2022년 12월 25일 key%%%%%%%%%%%%%%%%%%
{{
Especially, an interesting property of (holographic) entanglement entropy is that there is a concrete relation between the subsystem energy and its entanglement entropy in the limit that the system size is very much small, which is called the first law of entanglement entropy\cite{Bhattacharya:2012mi,Bianchi:2012ev,Nozaki:2013vta,Allahbakhshi:2013rda,Wong:2013gua,Momeni:2015vka,Park:2012lzs,Park:2015hcz,Kim:2016jwu,Jeong:2022zea}. In small subsystem limit, there is a relation, $\Delta E= \mathcal T \Delta S$, where $\Delta E$ is energy difference of the subsystem when it is excited and $\Delta S$ is the corresponding change of the entanglement entropy. $\mathcal T$ is called entanglement temperature. We note that it is widely discussed that the entanglement temperature is universally proportional to the invese of the subsystem size regardless of the shape and dimensionality of the entangling region.

%%%%%%%%%%%%%%%%%%%%%%%%%%%%%%%%%%%%%%%%%%%%%
%Our motivation is the entanglement entropies with different orientations enen if they share the same shape of the subsystem(or subset on the boundary space), they will show different value of the entanglment entropies due to appearance of vector order near the critical point.
}}

In this paper, we study Einstein SU(2) Yang-Mills theory 
%affected by an appearance of the vector order 
near critical point, 
by employing holographic entanglement entropy. In the dual field theory, a vector order appears and it breaks SO(3) rotational symmetry. Especially, we concentrate on some features of (holographic) entanglement entropy near critical point as follows. One may expect that entanglement entropy will perceive some of thermodynamic properties of field theory system near critical point. Since the gravity model undergoes second order phase transition near critical point, one may wonder if (holographic) entanglement entropy may show a scaling behavior like other quantities such as $\eta_{12}$, the ratio of shear viscosity to entropy density in anisotropic direction. 
{Another question is if entanglement entropy can provide a new order parameter like the vector order parameter that we discuss above. From this, one can recognize that the SO(3) spatial rotational symmetry breaks down to SO(2). 
}
Finally, we want to check if the first law of entanglement entropy near the critical point is still valid, keeping its universal properties of entanglement temperature, even though the SO(3) symmetry is broken by the vector order.

In the following, we will answer the questions that we raised above in order.
%The main result is twofold. We illustrate these main consquences below in order.
First of all, we compute entanglement entropies of subsystems on the boundary spacetime with shapes of ``wide and thin slabs'' and a ``cylinder''. The slabs are computable examples by applying analytic approaches\footnote{We note that there are some of earlier numerical works\cite{Arias:2012py,Cai:2012nm}, in which they discuss entanglement entropy of a slab in this background.} but for the cylinder case, we need numerics as well as analytic ones\footnote{We get analytic(algebraic) solutions of surface area for the cylinder on AdS boundary. To apply one of the boundary conditions to the surface, we use numerics. For the details, see Sec.\ref{Holographic entanglement calculation of cylinder}}. We study two different slabs, which share the same shape but we put them in different directions. More precisely, we consider a wide slab which is perpendicular to the vector order(the vector order is along $x_1-$direction) and another slab being parallel to the vector order. We call each of the entanglement entropy $S_1$, $S_2$ respectively. We define quantities, $\Delta_\varepsilon S_i\equiv S^{iso}_i - S^{aniso}_i\ (i=1,2)$, which shows how much the excess of entanglement entropy is when the boundary field theory system shows phase transition to anisotropic phase from isotropic phase. It turns out that $\Delta_\varepsilon S_i$ presents a scaling behavior such that
\begin{equation}
\Delta_\varepsilon S_i=\frac{2520\pi^2}{17\kappa_5^2}\Sigma_i \mathcal A_i^{(\varepsilon)}(d)T_c\left(1-\frac{T}{T_c}\right)^\beta, %\frac{1260\pi}{17},
\end{equation}
where the critical exponent $\beta$ turns out to be one, i.e. $\beta=1$. {$\Sigma_i$} is the cross sectional area of each slab, $\mathcal A^{(\varepsilon)}_i$ are the factors, showing ``$d$'' dependence, where $d$ is the thickness of the slabs and $T_c$ is the critical temperature. The leading behaviors of $\mathcal A^{(\varepsilon)}_i$ is given by
\begin{equation}
\mathcal A^{(\varepsilon)}_i(d) = \frac{281}{134400\pi^{7/2}}\Gamma\left(\frac{1}{3}\right)^3\Gamma\left(\frac{1}{6}\right)^3 d^2+O(d^4), 
\end{equation}
for both of $\mathcal A^{(\varepsilon)}_i$ but the next sub-leading is different from each other.
%where $\eta_i$ are numerical factors and $\eta_1\neq \eta_2$. 
For the subsystem with its shape of cylinder, we also compute the same quantity, $\Delta_\varepsilon S_{cy}=S^{iso}_{cy}-S^{aniso}_{cy}$ as we discussed above and find that 
\begin{equation}
\Delta_\varepsilon S_{cy}=\frac{5040\pi^2L_1}{17\kappa^2_5}\gamma^{(\varepsilon)}(a)T_c\left(1-\frac{T}{T_c}\right)^\beta,
\end{equation}
where 
\begin{equation}
\gamma^{(\varepsilon)}(a)=0.163313a^3+O(a^4),
\end{equation}
where $a$ and $L_1$ is the radius and the length of the cylinder and $\beta=1$. We note that to get this results, we utilize analytic as well as numerical methods. 

Second of all, once we set the cross-sectional area of the two slabs to be equal as $\Sigma\equiv \Sigma_1=\Sigma_2$,
% for the slabs, 
we can define an interesting new order parameter $\mathcal O_{12}\equiv S_1-S_2$ which vanishes in isotropic phase. However, once the dual field theory system gets into the anisotropic phase, it becomes
\begin{equation}
\mathcal O_{12}= S_1^{aniso}-S_2^{aniso}=-\frac{2520\pi^2}{17\kappa_5^2}\Sigma \mathcal A^{(\varepsilon)}(d)T_c\left(1-\frac{T}{T_c}\right)^\beta,
\end{equation}
%where
%\begin{equation}
%\alpha(d)=-\frac{2520\pi^2}{17\kappa_5^2}\Sigma \mathcal A^{(\varepsilon)}(d)T_c,
%\end{equation}
which also shows the same critical exponent $\beta=1$. %$d$ is the thickness of the slab and the leading behaviour of $\alpha (d)$ is proportional to $\sim d^4$. %This shows that condensation mediated by such a vector order is more influential to the entanglement entropy.
The leading behavior of $\mathcal A^{(\varepsilon)}(d)$ is proportional to $\sim d^4$.
More precisely, the leading behavior of {$\mathcal A^{(\varepsilon)} (d)$} is given by
\begin{equation}
\mathcal A^{(\varepsilon)}(d)=\frac{3 \sqrt{3}}{448 \pi ^{9/2}}\Gamma \left(\frac{1}{6}\right)^3 \Gamma \left(\frac{1}{3}\right)^3\ d^4+O(d^6).
\end{equation}

Finally, we study the first law of entanglement entropy in this framework. We find that even in the case that the vector order appears near critical point, the first law of entanglement entropy is still valid for both of the subsystems of the slabs and the cylinder. This means that subsystem energy and entanglement entropy are proportional to  each other and the ratio of one to another is the same $\mathcal T$, which is the entanglement temperature when the subsystem is out of the critical point. Especially, by employing analytic as well as numerical methods, we determine entanglement temperature of the subsystem with the shape of cylinder, which is given by
\begin{equation}
\mathcal T_{cy}=\frac{c_{ent}}{a},{\rm \ \ where\ \ }c_{ent}=0.163004\pm0.000001,
\end{equation} 
where  $a$ is the radius of the cylinder.

{%\color{blue}
%We close this section with a final remark. 
%The condensation appears in relatively {\color{red} {\it infrared}} region. In dual field theory, to see this, we need to make our subsystem size be increased. This corresponds that the minimal surface area extending to the bulk geometry probes deeper in the bulk spacetime and it means probing relatively {\color{red} {\it infrared} regions} in {\color{red} holographic} dictionary.
We close this section with a remark.
The facts that the critical behaviors and its critical exponent $\beta=1$ of the {entanglement} entropies that we compute are turned out to be universal features in our analytic approach. However, anisotropic features also appear in the factors, for example, $\mathcal A^{(\varepsilon)}_i(d)$ in thin and wide slab cases. 
In the small ``$d$'' region, we only see the leading behavior of entanglement entropy which is proportional to $d^2$. However, as $d$ increases, the subleading corrections become important and it shows spatial anisotropy and it may depend on an angle between the direction of the vector order and the axis that the slab is lying in.
 Our analysis manifestly shows that the leading behaviors of the anisotropy in the entanglement entropy is contained the coefficient, $\mathcal A^{(\varepsilon)}_i(d)$, which have information of directional dependency of degrees of freedom when the vector order appears.
}

\section{Holographic model}
%==== TEMP 4 ::: 1201.5605 Section 2.1 Equations ==== \\
In this section, we will review the holographic model for anisotropic super fluids defined on its conformal boundary. To illustrate the model, we mostly follow the papers\cite{Basu:2011tt,Oh:2012zu}. %We use the same symbols for the variables that appear in the discussion to avoid possible confusions. 
We begin with the holographic model given by
\begin{equation}
S=\int d^5x \sqrt{-G}\left(\frac{1}{\kappa^2_5}\left(R+\frac{12}{L^2}\right)-\frac{1}{4g^2}F^a_{MN}F^{aMN}\right),
\end{equation}
where $\kappa_5$ is 5-dimensional gravity constant, $g$ is the gauge coupling for $SU(2)$ gauge field $B_M^a$. $L$ is the length scale for cosmological constant and we set $L=1$ in the following discussion. The field strength for the gauge fields is given by
\begin{equation}
F^a_{MN}=\partial_MB^a_N-\partial_NB^a_M-\varepsilon ^{abc}B^b_MB^c_N,
\end{equation}
where indices with upper case Latin letters as $M$, $N$... are spacetime indices and the indices with the  lower case Latin letters are gauge indices, and they run as $a,b,c=1,2,3$. $\varepsilon^{abc}$ is fully anti-symmetric tensor.

By applying variational principle of the fields to the action, we get their equations of motion. They are
\begin{align}
W_{MN} &\equiv R_{MN}+4G_{MN}-\kappa^2_5\left(T_{MN}-\frac{1}{3}T^P_PG_{MN}\right)=0,\\
\label{YAng-mill-equation}
Y^{aN} &\equiv \nabla_MF^{aMN}-\varepsilon^{abc}B^b_MF^{cMN}=0,
\end{align}
where $W_{MN}$ is Einstein equation and $Y_M^a$ is gauge field equation. $T_{MN}$ is stress-energy tensor being given by
\begin{equation}
T_{MN}=\frac{1}{g^2}\left(F^a_{MP}F^{Pa}_N-\frac{1}{4}F_{PQa}F^{PQa}G_{MN}\right).
\end{equation}

Now, we are going to get their solutions. The forms of the solutions that we try are 
\begin{align}
B&=\phi(r)\tau^3dt+\omega(r)\tau^1dx_1,\\ 
\label{background-metruc}
ds^2&=-N(r)\sigma^2(r)dt^2+\frac{dr^2}{N(r)}+r^2f^{-4}(r)dx_1^2+r^2f^2(r)(dx_2^2+dx_3^2),
\end{align}
where $x_1$,$x_2$, and $x_3$ are the boundary spatial coordinates.
One of the solutions is 5-dimensional charged black brane solution, which is given by
\begin{align}
\phi(r)&={\mu}(1-\frac{1}{r^2}),\ \ \omega(r)=0, \\ \nonumber
\sigma(r)&=f(r)=1\ \textrm{and}\ N(r)=N_0(r)\equiv r^2-\frac{m}{r^2}+\frac{2{\mu}^2\alpha^2}{3r^4},
\end{align}
where $\mu$ is chemical potential for the gauge field $\phi$, $m=1+\frac{2\mu^2\alpha^2}{3}$ is the mass of the black brane and the constant $\alpha^2\equiv\frac{\kappa^2_5}{g^2}$. We note that the horizon of the black brane is located at $r=1$ in this solution by employing an appropriate coordinate rescaling such that $r \rightarrow \lambda r$ and $\{t,x_1,x_2,x_3\}\rightarrow \lambda^{-1}\{t,x_1,x_2,x_3\}$ with a real constant $\lambda$. In this rescaled coordinate, the chemical potential $\mu$ is dimensionless and it turns out that at $\mu=4$, the normalizable mode of solution $\omega (r)$ appears.

Now, we are interested in another kind of solutions, where $\omega(r)$ does not vanish. The way how to get a solution is to solve the gauge field equations(\ref{YAng-mill-equation}) by assuming that the $\omega(r)$ is non-zero but still small. It turns out that the form of the solution $\omega(r)$ is given by
\begin{equation}
\omega(r)=\varepsilon\frac{r^2}{(r^2+1)^2}+O(\varepsilon^2),
\end{equation}
where $\varepsilon$ is a small parameter representing the magnitude of $\omega(r)$. 

The solution $\omega(r)$ is $x_1-$directional gauge field, which breaks $SO(3)$ global rotation symmetry of the spacetime into $SO(2)$. This becomes more manifest when we compute back reactions to the black brane background spacetime. To compute backreactions, we assume the $\alpha^2$ is also parametrically small and so stress-energy tensor of the gauge field excitation, $\phi(r)$ and $\omega(r)$ does not significant change the background. We briefly list the results below considering backreactions to the field $\phi$ upto leading order in $\varepsilon^2$ and the background spacetime upto leading order in $\varepsilon^2\alpha^2$.
\begin{align}
\omega(r)&=\varepsilon\frac{r^2}{(r^2+1)^2}+O(\varepsilon^2),\\
\phi(r)&=4\left(1-\frac{1}{r^2}\right)+\frac{\varepsilon^2}{4}\left(\frac{1+2r^2}{3r^2(1+r^2)^3}-\frac{1}{8}+\frac{281}{1680}\left(1-\frac{1}{r^2}\right)\right)+O(\varepsilon^3)
\end{align}
and
\begin{align}
\label{metric-factor}
\sigma(r)&=1-\varepsilon^2\alpha^2\frac{2}{9(1+r^2)^3},\ f(r)=1-\varepsilon^2\alpha^2F(r)
\\ \nonumber
N(r)&=N_0(r)\left(1-\alpha^2N_\alpha(r)+\varepsilon^2\alpha^2N_\varepsilon(r)\right)
\end{align}
where 
\begin{equation}
\label{metric-f-factor2}
F(r)=\frac{1}{18}\frac{1-2r^2}{(1+r^2)^4},
\end{equation}
\begin{align}
\label{metric-N-factor2}
N_0(r)&=r^2-\frac{1}{r^2},\ N_\alpha=\frac{32}{3}\frac{1}{r^2(r^2+1)},\ \text{and}\ 
N_\varepsilon(r)=\frac{4}{9}\frac{1}{r^2}\left(\frac{281}{560}\frac{1}{r^2}-\frac{2+6r^2+3r^4}{2(1+r^2)^4}\right)
\end{align}
Sometimes, we employ the radial coordinate $z$, which is defined by $z=\frac{1}{r}$. 
In such a case, the metric is
\begin{equation}
ds^2=\frac{1}{z^2}\left\{-z^2N(z)\sigma^2(z)dt^2+\frac{dz^2}{z^2N(z)}+f^{-4}(z)dx_1^2+f^2(z)(dx_2^2+dx_3^2)\right\},
\end{equation}
where,
\begin{align}
\label{fzNzform}
f(z)&=1-\varepsilon^2\alpha^2F(z),
\\ \nonumber
N(z)&=N_0(z)\left(1-\alpha^2N_\alpha(z)-\varepsilon^2\alpha^2N_\varepsilon(z)\right).
\end{align}
and
\begin{align}
\label{Fz_in_fz_form}
F(z)=\frac{1}{18}\frac{z^6(z^2-2)}{(z^2+1)^4},
\end{align}
\begin{align}
\label{N0zNalphazNepsilonzform}
N_0(z)=\frac{1}{z^2}-z^2,\ N_\alpha(z)=\frac{32}{3}\frac{z^4}{1+z^2},\ \text{and}\ N_\varepsilon(z)=\frac{4}{9}z^4\left(\frac{281}{560}-\frac{2z^6+6z^4+3z^2}{2(1+z^2)^4}\right)
\end{align}
%\begin{align}
%\sigma (r)&=1-\varepsilon^2\alpha^2\frac{2}{9(1+r^2)^3},\ f(r)=1-\varepsilon^2\alpha^2\frac{(1-2r^2)}{18(1+r^2)^4}\\ \nonumber
%N(r)&= r^2 - \frac{1}{r^2}+\frac{32\alpha^2}{3}\left(\frac{1}{r^4}-\frac{1}{r^2}\right)-\varepsilon^2\alpha^2\frac{4}{9r^2}\left(\frac{1+2r^2}{r^2(1+r^2)^3}-\frac{3r^2}{2(1+r^2)^2}\right.
%\\ \nonumber
%&+\left.\frac{281}{560}\left(1-\frac{1}{r^2}\right)\right)
%\end{align}
%where the factor $\sigma(r)$ and $f(r)$

Finally, we discuss some of the black brane thermodynamics. The black brane temperature is given by
%\begin{equation}
%a(r)=a_0(r)+\varepsilon a_1(r)+\varepsilon^2 a_2(r)+\cdots
%\end{equation}
%\begin{equation}
%a_i(r)=a_{i,0}(r)+\alpha^2 a_{i,2}(r)+\alpha^4 a_{i,4}(r)+\cdots
%\end{equation}
\begin{equation}
\label{temp-bb}
T=\frac{1}{\pi}\left(1-\frac{16}{3}\alpha^2+\frac{17}{1260}\varepsilon^2\alpha^2\right)
\end{equation}
and the black brane entropy is 
\begin{equation}
S_{\rm black-brane}=\frac{2\pi}{\kappa^2_5}V_3,
\end{equation}
where {we take the horizon located at $z=1$} and $V_3$ is the spatial coordinate volume, $V_3=\int d^3 \vec x$. We note that the critical temperature, $T_c$ is given by
\begin{equation}
\label{temp-critical-bb}
T_c=\frac{1}{\pi}\left(1-\frac{16}{3}\alpha^2\right).
\end{equation}
%and therefore, the magnitude of the backreaction, $\varepsilon^2\alpha^2$ can be written as
%\begin{equation}
%\varepsilon^2\alpha^2=\frac{1260\pi}{17}(T-T_c).
%\end{equation}

%\begin{equation}
%ds^2=-N(r)\sigma^2(r)dt^2+\frac{dr^2}{N(r)}+r^2 f^{-4}(r)dx^2 + r^2 f^2(r)(dy^2+dz^2)
%\end{equation}

\section{Holographic computations of entanglement entropy of Wide and thin slabs}
In this section, we will discuss holographic entanglement entropy probes near critical point, $T=T_c$ in the presence of the vector mode $\omega(r)$ and considering its backreaction to the background geometry. %For the later use, we may define the following quantities.
%\begin{eqnarray}
%\label{f-form}
%f(r)\equiv1-\varepsilon^2\alpha^2F(r),%+O(\varepsilon^a\alpha^b), 
%\\ 
%\label{N-form}
%N(r)\equiv N_0(r)\left(1-\alpha^2N_\alpha-\varepsilon^2\alpha^2N_\varepsilon(r)\right)
%N(r)\equiv\left(r^2-\frac{1}{r^2}\right)\left(1-\alpha^2N_\alpha-\varepsilon^2\alpha^2N_\varepsilon(r)\right),%+O(\varepsilon^a\alpha^b),
%\end{eqnarray}
%where $a$ or $b>2$ 
%where
%\begin{align}
%F(r)&=\frac{(1-2r^2)}{18(1+r^2)^4}, \\
%N_0(r)&=r^2-\frac{1}{r^2},
%N_\alpha(r)= \frac{32}{3}\frac{1}{r^2(r^2+1)}, {\rm \ \ and \ \ }
%N_\varepsilon(r)= \frac{4}{9}\frac{1}{r^2}\left( \frac{281}{560}\frac{1}{r^2}-\frac{2+6r^2+3r^4}{2(1+r^2)^4}\right).
%\end{align}
%Sometimes, we employ the radial coordinate $z$, which is defined by $z=\frac{1}{r}$.
%\subsection{}
We consider a subsystem on AdS boundary with its shape of wide slab. There are two different ways to put the slab, which are to put that on $x_2-x_3$ plane and on $x_3-x_1$ plane(remember that $x_1$- direction is parallel to the vector order $\omega(r)$).

\subsection{The slab on $x_2-x_3$ plane} 
Think of a slab on AdS boundary where the slab is given by $-\frac{L_2}{2}\leq x_2\leq \frac{L_2}{2}$,  $-\frac{L_3}{2}\leq x_3\leq \frac{L_3}{2}$ and $-\frac{d}{2}\leq x_1\leq \frac{d}{2}$. We take $L_2$ and $L_3$ to be very large, which makes the subsystem have translational symmetric directions along $x_2$ and $x_3$ axes($L_2,L_3 \rightarrow \infty$). Now we compute a surface area in $d+1$-dimensional bulk from the slab on AdS boundary, which is given by
\begin{equation}
A_1 = \lim_{\delta\rightarrow 0}2L_2L_3\int^{\frac{1}{\delta}}_{r_*} dr\ r^3\sqrt{\frac{f^4(r)}{r^2N(r)}+\left(\frac{dx_1}{dr}\right)^2},
\end{equation}
where boundary of the surface, $A_1$(located at $r=\frac{1}{\delta}$ together with $\delta \rightarrow 0$) matches with the boundary of the slab on AdS boundary.
$A_1$ is hanged down into the bulk and there is a minimum value of the radial variable, $r$. We address that minimum value as $r_*$. %It is also given when $x=0$.
%where the $\varepsilon$ is a small constant to regularize the area since its near boundary behavior might be divergent.  
%where the $r_*$ is a small constant to regularize

Holographic entanglement entropy, $S$ is related to the area of surface in the bulk, $A$ as
\begin{equation}
S=\frac{2\pi}{\kappa^2}A,
\end{equation}
where $A$ needs to be minimized followed by Ryu-Takayanagi-prescription\cite{Ryu:2006bv,Ryu:2006ef}. To extremize the surface, we apply variational principle to the area and we get the condition of extremum being given as
\begin{equation}
\label{yz-equation}
\frac{d}{dr}\left(r^3\frac{x_1'}{\sqrt{(x_1')^2+\frac{f^4(r)}{r^2N(r)}}}\right)=0,
\end{equation}
where the prime denotes derivative with respect to $r$.
%To interpret the solution of the equation(\ref{yz-equation}), 
Let us close look at the inside the round bracket in (\ref{yz-equation}). The $A_1$ is hanged down deep into the bulk spacetime and it becomes deepest when $\frac{dr}{dx_1}=0$ and it is given when $x_1=0$. %We call the maximal depth of the coordinate $r$ as $r=r_*$ and it is given when $x_1=0$. 
The solution of equation(\ref{yz-equation}) is %given in such a way 
that the quantity inside of round bracket is a constant and the above argument fixes the constant as $r^3\frac{x_1'}{\sqrt{(x_1')^2+\frac{f^4(r)}{r^2N(r)}}}=r_*^3$.

With such an identification, the solution is able to be written as
\begin{equation}
x_1'=\pm\frac{r_*^3f^2(r)}{r^4\sqrt{\left(1-\frac{r_*^6}{r^6}\right)N(r)}},
\end{equation}
and we substitute the solution into the surface area $A_1$ to remove $\frac{dx_1}{dr}$ in it. Then, $A_1$ is given by
\begin{equation}
\label{A1-area-formula}
A_1=2L_2L_3 \lim_{\delta\rightarrow 0}\int^{\frac{1}{\delta}}_{r_*} dr\ \frac{r^2f^2(r)}{\sqrt{N(r)}}\left(1-\frac{r^6_*}{r^6}\right)^{-\frac{1}{2}}.
%\sqrt{\frac{f^4(r)}{r^2N(r)}+\left(\frac{dx}{dr}\right)^2}
\end{equation}
%where $c_x$ is an integration conatant.
%Near $r\rightarrow\infty$, $x\rightarrow\pm d/2$.

%\begin{equation}
%x^\prime(r)=\frac{1}{c_x  r^5}+\frac{\frac{281 \alpha ^2 \varepsilon ^2}{1260}+\frac{32 \alpha ^2}{3}+1}{2 c_x r^9}+\frac{\frac{2 \alpha ^2 \varepsilon
%   ^2}{9}+\frac{1}{2} \left(-\frac{1121 \alpha ^2 \varepsilon ^2}{1260}-\frac{32 \alpha ^2}{3}+\frac{1}{c_x ^2}\right)}{c_x 
%   r^{11}}+O\left(\left(\frac{1}{r}\right)^{12}\right)
%\end{equation}
%\begin{equation}
%x=\pm\left(\frac{d}{2}-\frac{1}{4c_xr^4}+\cdots\right)
%\end{equation}
%\begin{equation}
%\frac{A_x}{L_yL_z}=\frac{r^2}{2}+\frac{\alpha ^2 \left(-\frac{281 \varepsilon ^2}{5040}-\frac{8}{3}\right)-\frac{1}{4}}{r^2}+\frac{\frac{\alpha ^2 \left(187 \varepsilon
%   ^2+4480\right)}{3360}-\frac{1}{8 c_x^2}}{r^4}+O\left(\left(\frac{1}{r}\right)^{11}\right)
%\end{equation}
Moreover, between $r_*$ and $d$(the thickness of the slab lying in the $x_1$-direction), there is the following relationship:
\begin{equation}
\label{d-rstar-relation-a1}
d=2\lim_{\delta\rightarrow 0}\int^{\frac{1}{\delta}}_{r_*}\frac{dr}{\sqrt{\frac{r^2N(r)}{f^4(r)}\left(\frac{r^6}{r_*^6}-1\right)}},
\end{equation}
which can be easily derived by using an identity, $\int^{d/2}_{-d/2}dx_1=2\lim_{\delta\rightarrow 0}\int^{\frac{1}{\delta}}_{r_*}\frac{dr(x_1)}{r^\prime(x_1)}$, where the prime denotes that the derivative with respect to its argument. 

To evaluate the integration for the surface area, $A_1$ given in (\ref{A1-area-formula}), we need to look at the metric factors, $N(r)$ and $f(r)$ in (\ref{metric-factor}) carefully. The metric is obtained by taking into account backreactions from the vector order $\omega(r)$ upto its leading order corrections. The leading correction is order of $\varepsilon^2 \alpha^2$. We note again that $\alpha^2=\frac{\kappa^2_5}{g^2}$ is regarded as small parameter as well as $\varepsilon^2 \ll 1$ and we deal with those perturbatively. In sum, we expand the surface area $A_1$ upto leading order correction in $\varepsilon^2\alpha^2$ and evaluate the integrations. Then, we have $\alpha^2$ and $\varepsilon^2$ corrections together with the zeroth order terms in $\alpha$ and $\varepsilon$ in $A_1$. Each integration is not so easy to get an analytic and compact form, so we assume that $d$ is small as well as $r_* \gg 1$ to get a series form of the integrands in small $d$. Firstly, by applying all the arguments that we address above, we evaluate the relation(\ref{d-rstar-relation-a1}), which is given by
%yz-plane (equation 4)
\begin{equation}
d=\frac{2 \sqrt{\pi }\ \Gamma \left(\frac{2}{3}\right)}{\Gamma \left(\frac{1}{6}\right)}\frac{1}{r_*}
+\frac{\sqrt{\pi }\ \Gamma \left(\frac{4}{3}\right) \left(1260 +13440\alpha ^2 + 281\alpha ^2 \varepsilon ^2 \right)}{7560\ \Gamma \left(\frac{11}{6}\right)} \left(\frac{1}{r_*}\right)^5
-\frac{\sqrt{\pi }\ \Gamma \left(\frac{5}{3}\right) \left(4480\alpha ^2+187\alpha ^2 \varepsilon ^2\right)}{2520\ \Gamma \left(\frac{13}{6}\right)} \left(\frac{1}{r_*}\right)^7...,
\end{equation}
and its inverse relation is 
\begin{equation}
\frac{1}{r_*} =
\frac{\Gamma \left(\frac{1}{6}\right)}{2 \sqrt{\pi }\ \Gamma \left(\frac{2}{3}\right)}\ d
-\frac{\Gamma \left(\frac{1}{6}\right)^6 \Gamma \left(\frac{4}{3}\right) \left(1260+13440 \alpha ^2+281 \alpha ^2 \varepsilon ^2\right)}{483840\ \pi^{5/2}\ \Gamma \left(\frac{2}{3}\right)^6 \Gamma \left(\frac{11}{6}\right)}\ d^5 
+\frac{\Gamma \left(\frac{1}{6}\right)^8 \Gamma \left(\frac{5}{3}\right) \left(4480\alpha ^2+187\alpha ^2 \varepsilon ^2\right)}{645120\ \pi ^{7/2} \Gamma \left(\frac{2}{3}\right)^8 \Gamma   \left(\frac{13}{6}\right)}\ d^7...
\end{equation}
Now, we evaluate the minimal surface area $A_1$ by replacing $r_*$ in it with $d$
by using the above relation together.% with the form of the minimal surface area $A_1$, we get
%it manipulation request

In fact, $A_1$ contains divergence which depends on the radial cut-off, $\frac{1}{\delta}$ near AdS boundary. This divergence already appears in entanglement entropy computation in pure AdS background, and we call that $S_1^{(AdS)}$. Therefore, we need to define a renormalized entanglement entropy, where we define the renormalized version of entanglement entropy as
\begin{equation}
S_1^{(ren)}=S_1-S^{(AdS)}.
\end{equation}

It turns out that the entanglement entropy in AdS background is given by
\begin{equation}
S^{(AdS)}=\frac{2\pi}{\kappa^2_5}\Sigma_1A^{(AdS)},
\end{equation}
where $\Sigma_1$ is the coordinate volume of the slab,
\begin{equation}
\Sigma_1=L_2L_3.
\end{equation}
The $A^{(AdS)}$ is given by
\begin{equation}
A^{(AdS)}=\lim_{\delta\rightarrow0}\frac{1}{\delta^2}-\frac{32 \pi ^{9/2}}{3 \sqrt{3} \Gamma \left(\frac{1}{3}\right)^3 \Gamma\left(\frac{1}{6}\right)^3}\ d^{-2},
\end{equation}
which contains the divergence appearing near AdS boundary and the finite term being proportional to $d^{-2}$. Now, we will subtract this quantity from $S_1$, and we get the renormalized entanglement entropy, which is given by
%We get $S_1$ as
\begin{equation}
S^{(ren)}_1=\frac{2\pi}{\kappa^2_5}\Sigma_1 \mathcal A_1,
\end{equation}
%where $\Sigma_1$ is the coordinate volume of the slab,
where $\mathcal A_1$ is 
%====   TEMP 5a::: Slab Area per LyLz   ====\\
\begin{equation}
\mathcal A_1 =\mathcal A_1^{(0)}+\alpha^2\mathcal A_1^{(\alpha)}+\varepsilon^2\alpha^2\mathcal A_1^{(\varepsilon)},
%\frac{1}{\delta^2}+C^{-2}_{yz}\ d^{-2}+C^2_{yz}\ d^2+C^4_{yz}\ d^4+C^6_{yz}\ d^6+O(d^7),
\end{equation}
and
\begin{align}
\label{a1-zero-area}
\mathcal A^{(0)}_1  &=\frac{3}{320\pi ^{7/2}} \Gamma \left(\frac{1}{3}\right)^3 \Gamma \left(\frac{1}{6}\right)^3\ d^2\\ \nonumber
&+ \frac{27}{16384\pi^9}\Gamma\left(\frac{1}{3}\right)^6\Gamma\left(\frac{1}{6}\right)^6\left(1-\frac{13\sqrt{3}}{1800\pi^{5/2}}\Gamma\left(\frac{1}{3}\right)^3\Gamma\left(\frac{1}{6}\right)^3\right)\ d^6+O(d^7),
\\
\label{a1-alpha-area}
\mathcal A^{(\alpha)}_1 &=\frac{1}{10\pi ^{7/2}} \Gamma \left(\frac{1}{3}\right)^3 \Gamma \left(\frac{1}{6}\right)^3\ d^2
-\frac{3\sqrt{3}}{28 \pi ^{9/2}} \Gamma \left(\frac{1}{3}\right)^3 \Gamma
   \left(\frac{1}{6}\right)^3\ d^4\\ \nonumber
&+\frac{9}{256\pi^9}\Gamma\left(\frac{1}{3}\right)^6\Gamma\left(\frac{1}{6}\right)^6\left(1-\frac{13\sqrt{3}}{1800\pi^{5/2}}\Gamma\left(\frac{1}{3}\right)^3\Gamma\left(\frac{1}{6}\right)^3\right)\ d^6+O(d^7),
\\
\label{a1-epsilon-area}
\mathcal A^{(\varepsilon)}_1 &=\frac{281}{134400\pi ^{7/2}} \Gamma \left(\frac{1}{3}\right)^3 \Gamma \left(\frac{1}{6}\right)^3\ d^2-\frac{561\sqrt{3}}{125440 \pi ^{9/2}} \Gamma \left(\frac{1}{3}\right)^3 \Gamma
   \left(\frac{1}{6}\right)^3\ d^4\\ \nonumber
&-\frac{837}{1146880\pi^9}\Gamma\left(\frac{1}{3}\right)^6\Gamma\left(\frac{1}{6}\right)^6\left(1+\frac{3653\sqrt{3}}{502200\pi^{5/2}}\Gamma\left(\frac{1}{3}\right)^3\Gamma\left(\frac{1}{6}\right)^3\right)\ d^6+O(d^7).
\end{align}

\subsection{The slab on $x_1-x_3$ plane}
To put the slab on the $x_1-x_3$ plane, we parametrize the slab as
$-\frac{L_3}{2}\leq x_3\leq \frac{L_3}{2}$,  $-\frac{L_1}{2}\leq x_1\leq \frac{L_1}{2}$ and $-\frac{d}{2}\leq x_2\leq \frac{d}{2}$. Again, we take $L_1$ and $L_3$ to be very large, meaning that we take a parametric limit as $L_1$ and $L_3\rightarrow\infty$. The formula for the surface area hanged down in the bulk is given by
\begin{equation}
\label{surface-area_A2222}
A_2 = L_1L_3\lim_{\delta\rightarrow0}\int^{1/\delta}_{r_\#}dr\ r^3\sqrt{\frac{1}{r^2f^2(r)N(r)}+\left(\frac{dx_2}{dr}\right)^2},
\end{equation}
and its equation for extremum is
\begin{equation}
\frac{d}{dr}\left(r^3\frac{x_2'}{\sqrt{(x_2')^2+\frac{1}{r^2f^2(r)N(r)}}}\right)=0,
\end{equation}
where again the boundary of the surface area, $A_2$ is coincide with the boundary of the slab at $r=\frac{1}{\delta}$ where we take a limit of $\delta\rightarrow0$.
We also define $r_\#$ this time, which is the minimum of the $r$-value. %(we note that even though we use the same symbol, $r_*$ with the case of slab lying on $x_1-x_2$ plane, you see the relation between $d$ and $r_$ is different from that case as addressed in (\ref{d-r*-relation}). Th). 
By applying the similar argument that we addressed in the case with slab lying on $x_1-x_2$ plane, we get
%and its solution is given by
\begin{equation}
\label{solution-of-ex-A2}
x_2'=\pm\frac{r^3_\#}{r^4f(r)\sqrt{\left(1-\frac{r^6_\#}{r^6}\right)N(r)}}.
\end{equation}
Now, we plug the solution of the equation of extremum(\ref{solution-of-ex-A2}) into the expression of the surface area $A_2$(\ref{surface-area_A2222}), we get
\begin{equation}
\label{A2-exexepress}
A_2=2L_1L_3\lim_{\delta\rightarrow0}\int^{\frac{1}{\delta}}_{r_\#} dr\ \frac{r^2}{f(r)\sqrt{N(r)}}\left(1-\frac{r^6_\#}{r^6}\right)^{-\frac{1}{2}}.
%\sqrt{\frac{f^4(r)}{r^2N(r)}+\left(\frac{dx}{dr}\right)^2}
\end{equation}
%where $c_y$ is an integration conatant.
%\begin{equation}
%y^\prime=\frac{c_y}{r^5}+\frac{c_y \left(\frac{281 \alpha ^2 \varepsilon ^2}{1260}+\frac{32 \alpha ^2}{3}+1\right)}{2 r^9}+\frac{c_y
%   \left(\frac{1}{2} \left(-\frac{1121 \alpha ^2 \varepsilon ^2}{1260}-\frac{32 \alpha ^2}{3}+c_y^2\right)-\frac{\alpha ^2 \varepsilon
%   ^2}{9}\right)}{r^{11}}+O\left(\left(\frac{1}{r}\right)^{12}\right)
%\end{equation}

%\begin{equation}
%\frac{A_y}{L_xL_z}=\frac{r^2}{2}+\frac{\alpha ^2 \left(-\frac{281 \varepsilon ^2}{5040}-\frac{8}{3}\right)-\frac{1}{4}}{r^2}+\frac{\frac{\alpha ^2 \left(467 \varepsilon
%   ^2+4480\right)}{3360}-\frac{c_y^2}{8}}{r^4}+O\left(\left(\frac{1}{r}\right)^{11}\right)
%\end{equation}

%By using the similar method, w
We also get the relation between the minimum value of $r$, $r_\#$ and $d$ in this case too. The form of the expression is
\begin{equation}
d=2\lim_{\delta\rightarrow0}\int^{\frac{1}{\delta}}_{r_\#}\frac{dr}{\sqrt{{r^2N(r)f^2(r)}{}\left(\frac{r^6}{r_\#^6}-1\right)}}.
\end{equation}
and its final form after performing the integration in it is given by
%====TEMP 3::: d and 1/r* in yz- and zx- plane====\\
%zx-plane (equation 10)
\begin{equation}
\label{d-r*-relation}
d=\frac{2 \sqrt{\pi }\ \Gamma \left(\frac{2}{3}\right)}{\Gamma \left(\frac{1}{6}\right)}\frac{1}{r_\#}
+\frac{\sqrt{\pi }\ \Gamma \left(\frac{4}{3}\right) \left(1260 +13440\alpha ^2 + 281\alpha ^2 \varepsilon ^2 \right)}{7560\ \Gamma \left(\frac{11}{6}\right)} \left(\frac{1}{r_\#}\right)^5
-\frac{\sqrt{\pi }\ \Gamma \left(\frac{5}{3}\right) \left(4480\alpha ^2+467\alpha ^2 \varepsilon ^2\right)}{2520\ \Gamma \left(\frac{13}{6}\right)} \left(\frac{1}{r_\#}\right)^7...
\end{equation}
The inverse relation of this is 
\begin{equation}
\label{d-r*-relationii}
\frac{1}{r_\#} =
\frac{\Gamma \left(\frac{1}{6}\right)}{2 \sqrt{\pi }\ \Gamma \left(\frac{2}{3}\right)}\ d
-\frac{\Gamma \left(\frac{1}{6}\right)^6 \Gamma \left(\frac{4}{3}\right) \left(1260+13440 \alpha ^2+281 \alpha ^2 \varepsilon ^2\right)}{483840\ \pi^{5/2}\ \Gamma \left(\frac{2}{3}\right)^6 \Gamma \left(\frac{11}{6}\right)}\ d^5 
+\frac{\Gamma \left(\frac{1}{6}\right)^8 \Gamma \left(\frac{5}{3}\right) \left(4480\alpha ^2+467\alpha ^2 \varepsilon ^2\right)}{645120\ \pi ^{7/2} \Gamma \left(\frac{2}{3}\right)^8 \Gamma\left(\frac{13}{6}\right)}\ d^7...
\end{equation}

Together with the relation (\ref{d-r*-relationii}) and the expression of $A_2$ given in (\ref{A2-exexepress}), the holographic entanglement entropy %with extremum of the stretched surface  
is given by
\begin{equation}
S^{(ren)}_2=\frac{2\pi}{\kappa^2_5}\Sigma_2\mathcal A_2,
\end{equation}
where $S_2^{(ren)}=S_2-S^{(AdS)}$ and $\Sigma_2$ represents $\Sigma_2=L_1L_3$. We note that  again to evaluate the integration in the expression, $A_2$, we expand that upto leading order in $\alpha^2$ and $\varepsilon^2\alpha^2$ together with $\frac{1}{r_\#}$ expansion with an assumption that $r_\#\gg1$ and $d\ll1$.

The $\mathcal A_2$ is also defined as the similar fashion as we did in the previous computation, being given by
\begin{equation}
\mathcal A_2=\mathcal A^{(0)}_2+\alpha^2\mathcal A_2^{(\alpha)}+\varepsilon^2\alpha^2\mathcal A_2^{(\varepsilon)},
\end{equation}
where
\begin{eqnarray}
\label{a2-0-alpha-area}
\mathcal A^{(0)}_2=\mathcal A^{(0)}_1, {\ \ }\mathcal A_2^{(\alpha)}=\mathcal A_1^{(\alpha)}
\end{eqnarray}
and
\begin{align}
\mathcal A^{(\varepsilon)}_2&=\frac{281}{134400\pi ^{7/2}} \Gamma \left(\frac{1}{3}\right)^3 \Gamma \left(\frac{1}{6}\right)^3\ d^2 -\frac{1401\sqrt{3}}{125440 \pi ^{9/2}} \Gamma \left(\frac{1}{3}\right)^3 \Gamma\left(\frac{1}{6}\right)^3 d^4
\\ \nonumber
&-\frac{6723}{1146880\pi^9}\Gamma\left(\frac{1}{3}\right)^6\Gamma\left(\frac{1}{6}\right)^6\left(1+\frac{3653\sqrt{3}}{4033800\pi^{5/2}}\Gamma\left(\frac{1}{3}\right)^3\Gamma\left(\frac{1}{6}\right)^3\right) d^6+O(d^7).
\end{align}

%====   TEMP 5b::: Slab Area per LzLx   ====\\
%\begin{equation}
%\frac{A_y}{L_zL_x}=\frac{1}{\delta^2}+C^{-2}_{zx}\ d^{-2}+C^2_{zx}\ d^2+C^4_{zx}\ d^4+C^6_{zx}\ d^6+O(d^7)
%\end{equation}
%\begin{align}
%C^{-2}_{zx}&=C^{-2}_{yz}\\
%C^2_{zx}&=C^2_{yz}\\
%C^4_{zx}&=-\frac{3\sqrt{3}}{28 \pi ^{9/2}} \Gamma \left(\frac{1}{3}\right)^3 \Gamma
  % \left(\frac{1}{6}\right)^3 \left(\alpha^2+\frac{467}{4480}\varepsilon ^2\alpha ^2\right)\\
%C^6_{zx}&=\frac{27}{16384\pi^9}\Gamma\left(\frac{1}{3}\right)^6\Gamma\left(\frac{1}{6}\right)^6\left(1-\frac{13\sqrt{3}}{1800\pi^{5/2}}\Gamma\left(\frac{1}{3}\right)^3\Gamma\left(\frac{1}{6}\right)^3\right)\\ \nonumber
%&+\frac{9}{256\pi^9}\Gamma\left(\frac{1}{3}\right)^6\Gamma\left(\frac{1}{6}\right)^6\left(1-\frac{13\sqrt{3}}{1800\pi^{5/2}}\Gamma\left(\frac{1}{3}\right)^3\Gamma\left(\frac{1}{6}\right)^3\right)\alpha^2\\
%&-\frac{6723}{1146880\pi^9}\Gamma\left(\frac{1}{3}\right)^6\Gamma\left(\frac{1}{6}\right)^6\left(1+\frac{3653\sqrt{3}}{4033800\pi^{5/2}}\Gamma\left(\frac{1}{3}\right)^3\Gamma\left(\frac{1}{6}\right)^3\right)\varepsilon^2\alpha^2
%\end{align}
\subsection{Properties of holographic entanglement entropy near critical point}
\paragraph{Scaling behavior of entanglement entropy near critical point}
As we discussed in Sec.\ref{Introduction}, this Einstein-SU(2)Yang-Mills system undergoes second order phase transition near the critical point, $T=T_c$, which is the phase transition from ``isotropic phase'' to ``anisotropic phase'' being affected by the appearance of vector order $\omega(r)$. Again, the $\omega(r)$ is the spatial component, $B^{1}_{1}$ of Yang-Mills fields, $B^a_\mu$. %component of the Yang-Mills gauge fields. 

Now, we define the ``isotropic'' and the ``anisotropic'' phases as follows.
In isotropic phase, $\omega(r)=0$ and there is no backreaction to the background geometry from it.
Since there is no spatial component of Yang-Mills fields turned on, the $SO(3)$ isometry(spatial rotation symmetry) in the bulk spacetime is retained. However, once the field $\omega(r)$ is turned on, this spatial isometry is broken down to $SO(2)$ and we call it anisotropic phase. The field $\omega(r)$ is normalizable mode in the bulk, which means that this mode corresponds to a state in dual field theory. We note that $SO(3)$ isometry is broken spontaneously.

In many literatures\cite{Basu:2011tt,Oh:2012zu,Park:2016wch}, it is widely discussed that once $\omega(r)$ appears near critical point, the anisotropic phase is thermodynamically favored than the isotropic phase, and so there will be thermodynamic phase transition from isotropic phase to anisotropic phase, near the critical point.  

The leading order backreaction from the field solution, $\omega(r)$ 
%%%%%%%%%%%%%%%%%%%%%%%%%2022년 10월 3일%%%%%%%%%%%%%%
is order of $\varepsilon^2 \alpha^2$(Remember that $\omega(r)\sim\varepsilon $). Therefore, if we compute a quantity where we turn off $\varepsilon=0$, then it corresponds to the quantity in isotropic phase whereas if we turn on $\varepsilon$ and keep the leading order backreactions to the background geometry upto $\varepsilon^2 \alpha^2$, then that quantity will be that in anisotropic phase. We call them $Q^{iso}$ and $Q^{aniso}$ respectively for some quantity, $Q$.

Now, we want see how much entanglement entropy excess arises, when the system undergoes phase transition. To see this, we define
\begin{equation}
\Delta_{\varepsilon} S_i\equiv S_i^{iso}-S_i^{aniso}=S_i^{(ren)iso}-S_i^{(ren)aniso},
\end{equation}
where $i=1,2$ to express, $S_1$ and $S_2$ respectively. We note that $\Delta_{\varepsilon} S_i$ is order of $\varepsilon^2 \alpha^2 \sim \left(1-\frac{T}{T_c}\right)$, which will show scaling behavior near the critical temperature $T=T_c$. In fact, by using black brane (critical)temperature(\ref{temp-bb}) and (\ref{temp-critical-bb}), we get
%Therefore,
\begin{equation}
\label{resultyzcritical}
\Delta_\varepsilon S_i=\frac{2520\pi^2}{17\kappa_5^2}\Sigma_i \mathcal A_i^{(\varepsilon)}T_c\left(1-\frac{T}{T_c}\right)^\beta,
\end{equation}
where $\beta$ is the critical exponent of the entropy, and it turns out that 
\begin{equation}
\beta=1,
\end{equation}
upto leading order corrections in $\alpha$ and $\varepsilon$ in our analytic calculation.

%In this case, we also discuss the difference between the holographic entanglement entropies of anisotropic and isotropic phases as
%\begin{equation}
%\Delta S_2\equiv S_2^{aniso}-S_2^{iso}=\frac{2\pi}{\kappa_5^2}\Sigma_2 \mathcal A_2^{(\varepsilon)}(T-T_c)^\beta \frac{1260\pi}{17},
%\end{equation}
%where we also have the same critical exponent of $\beta=1$ but different coefficient due to there directional dependence from (\ref{resultyzcritical}).

\paragraph{Order parameter measuring spatial anisotropy from entanglement entropy}
%====== TEMP 6 ======
In fact, one can introduce a new order parameter near the critical point by employing (holographic) entanglement entropy. We define an order parameter of anisotropy near the critical point, being given by
\begin{equation}
O_{12}\equiv S_1-S_2,
\end{equation}
which denotes the difference of entanglement entropies between the slabs being perpendicular to the vector order, $x_1$-direction and lying along the vector order. Once we make the cross section area of the slabs be the same, $\Sigma_1=\Sigma_2\equiv \Sigma$, then the quantity $\mathcal O_{12}$ vanishes in isotropic phase. %This is because that the entanglement entropy shows no orientation dependence at all in isotropic phase.
%\begin{equation}
%\mathcal O_{12}\equiv S_1^{aniso}-S_2^{aniso}=\frac{2\pi}{\kappa_5^2}(\Sigma_1 \mathcal A_1^{(\varepsilon)}-\Sigma_2 \mathcal A_2^{(\varepsilon)})(T-T_c)^\beta \frac{1260\pi}{17},
%\end{equation}
 However, near the critical point, this quantity shows non-zero value such that
\begin{equation}
\mathcal O_{12}\equiv S_1^{aniso}-S_2^{aniso}=-\frac{2520\pi^2}{17\kappa_5^2}\Sigma  \mathcal A^{(\varepsilon)}T_c\left(1-\frac{T}{T_c}\right)^\beta, %\frac{1260\pi}{17},
\end{equation}
where
\begin{align}
\label{O1212}
\mathcal A^{(\varepsilon)}=\mathcal A^{(\varepsilon)}_1 - \mathcal A^{(\varepsilon)}_2&=\frac{3 \sqrt{3}}{448 \pi ^{9/2}}\Gamma \left(\frac{1}{6}\right)^3 \Gamma \left(\frac{1}{3}\right)^3\ d^4
+\frac{2943}{573440 \pi ^9}\Gamma \left(\frac{1}{6}\right)^6 \Gamma \left(\frac{1}{3}\right)^6\ d^6+O(d^7).
\end{align}
Again the critical exponent, $\beta=1$ in our analytic analysis.

The leading dependence on the thickness of the slab, ``$d$''in the order parameter, $\mathcal O_{12}$ is $\sim d^4$. The difference between $S_1$ and $S_2$ near the critical point stems from the $F(r)$ in the metric factor $f(r)$ which are given in (\ref{metric-factor}) and (\ref{metric-f-factor2}). The reason why this happens is that $g_{11}\neq g_{22}=g_{33}$, which are the spatial components of bulk spacetime metric factors(See the metric(\ref{background-metruc}). The leading backreaction to the metric from the vector order, appears at $N_{\varepsilon}(r)$ in the metric factor $N(r)$, but $N_{\varepsilon}(r)$ does not give spatial anisotropy in the bulk spacetime metric.
In fact, $N_{\varepsilon}(r)\sim O(r^{-4})$, whereas the metric factor $F(r)\sim O(r^{-6})$. Therefore, $F(r)$ gives subleading correction as contrasted with $N_{\varepsilon}(r)$ correction once we consider near AdS boundary expansion order by order in small $1/r_*$ or $1/r_\#$. However, Once we think of $\mathcal O_{12}=S_1-S_2$, then contributions from $F(r)$ becomes leading, it grows as the subsystem becomes larger being proportional to $d^4$ and also shows scaling behavior as we addressed in (\ref{O1212}).

\paragraph{The first law of entanglement entropy}
It is widely discussed that in small ``$d$'' region, there is an interesting relation between the energy and its entanglement entropy of the subsystem\cite{Bhattacharya:2012mi, Jeong:2022zea}. The relation is called the first law of entanglement entropy, which can be understood as an analogy of thermodynamic first law. In short, the relation is given by
\begin{equation}
\Delta E= \mathcal T \Delta S,
\end{equation}
where $\mathcal T$ is called entanglement temperature, which is inversely proportional to the subsystem size. %We note that $\mathcal T$ is not an usual termperture and so it is not the black brane term. 
$\Delta S$ means entanglement entropy changes from its background to a new state, where the background is pure AdS space, which corresponds to vacuum defined in (AdS)boundary field theory in holographic dictionary. An interesting property of the entanglement temperature is that it is universal in a sense that it only depends on the shape of the subsystem and the dimensionality of (AdS) boundary spacetime, not any other details of the subsystem.

We find that even in the case that the vector order appears, the universality does not be broken. To discuss this, let us see the energy of the subsystem when we consider wide and thin slabs defined in AdS boundary. In the following discussion, we restrict ourselves in the case of the subsystem in $4-$dimensional spacetime.  According to the the prescription suggested in \cite{Bhattacharya:2012mi,Hertzberg:2010uv,Huerta:2011qi,Lewkowycz:2012qr,Rosenhaus:2014woa,Rosenhaus:2014zza,Park:2015dia}, 
\begin{equation}
\Delta E=E-E_g=\int_{\rm subsystem} d^3x \langle T_{tt}\rangle,
\end{equation}
where $\langle T_{tt}\rangle$ is temporal component of boundary stress-energy tensor, the energy density of the subsystem when the subsystem is excited from its ground state. $\int_{\rm subsystem} d^3x$ is the spatial volume integration of the subsystem. When, the energy density $\langle T_{tt}\rangle$ is a constant in the subsystem, 
\begin{equation}
\Delta E_i=\Sigma_i d \langle T_{tt}\rangle,
\end{equation}
for each slab, where $\Delta E_1$ is the energy for the slab lying on $x_2-x_3$ plane whereas $\Delta E_2$ is the energy for the slab lying on $x_1-x_3$ plane.
$E$ is the subsystem energy, and $E_g$ is its ground state energy, which corresponds to pure AdS space in its gravity dual. $\langle T_{tt}\rangle$ is related to black brane mass and the relation is
\begin{equation}
\langle T_{tt}\rangle=\frac{3M}{2\kappa^2_5},
\footnote{In general, $\langle T_{tt}\rangle = \frac{(d-1)M}{2\kappa^2_{d+1}}$, where $d$ is the dimensionality of AdS boundary spacetime\cite{Bhattacharya:2012mi}.}
\end{equation}
where $M$ is mass of the 5-dimensional charged black brane near critical point. We read off $M$ from the coefficient of $1/r^4$ term in the metric factor, $N(r)/r^2$ in its large $r$(small $1/r$) expansion.

In isotropic phase, the mass of the charged black brane, $M$ is given by
\begin{equation}
\label{MASS-subsystem}
M^{iso}=1+\frac{32}{3}\alpha^2.%+ \frac{281}{1260}\varepsilon^2 \alpha^2.
\end{equation}
We note that we rescale the coordinate variables, $r$, $t$ and $x^i$ in the metric to fix the the charged black brane horizon $r_h=1$.

The first term in (\ref{MASS-subsystem}) is related to the energy difference between black brane and pure AdS space. Once we turn on the temporal component of Yang-Mills fields, $B^3_0=\phi(r)$, the black brane becomes charged and so (electric) chemical potential comes in. This effect comes with $\alpha^2$ corrections in the spacetime metric, and so in $M$. %Near the critical point, once we have the normalizable mode of $B^1_1=\omega(r)$, the the system undergoes thermodynamic phase transition and gets into anisotropic phase. The energy change will come into the system with $\epsilon^2\alpha^2$ corrections, which is the last term in (\ref{MASS-subsystem}). 
In summary, we have
\begin{equation}
\Delta E_i=  \Delta_0 E_i+\Delta_\alpha E_i,
\end{equation}
where
\begin{equation}
\Delta_0 E_i= \frac{3}{2\kappa^2_5}\Sigma_i d{\rm \ \ and \ \ }\Delta_\alpha E_i= \frac{16}{\kappa^2_5}\alpha^2\Sigma_i d.
\end{equation}
We note that $\Delta_0 Q$ represents the difference between the quantity, $Q$ computed in black brane background and in pure AdS background.  $\Delta_\alpha Q$ is the $\alpha^2$ correction to the quantity, $Q$ when the chemical potential is turned on.
The entanglement entropy change is given by
\begin{equation}
\Delta S_i=\Delta_0 S+\Delta_\alpha S_i,
\end{equation}
where
\begin{equation}
\Delta_0 S_i=\frac{2}{\kappa^2_5}\Sigma_i \mathcal A^{(0)}_i {\rm \ \ and\ \ }
\Delta_\alpha S_i=\frac{2}{\kappa^2_5}\alpha^2\Sigma_i \mathcal A^{(\alpha)}_i,
\end{equation}
and the surface area $\mathcal A^{(0)}_i$ and $\mathcal A^{(\alpha)}_i$ are given in (\ref{a1-zero-area}), (\ref{a1-alpha-area}), and (\ref{a2-0-alpha-area}). By considering all the details in the above discussion, we get
\begin{equation}
\label{definition_of_entanglement_entropy_temperature}
\lim_{d\rightarrow0}\frac{\Delta E_i}{\Delta S_i}=\lim_{d\rightarrow0}\frac{\Delta_0 E_i}{\Delta_0 S_i}=\lim_{d\rightarrow0}\frac{\Delta_\alpha E_i}{\Delta_\alpha S_i}=\mathcal T,
\end{equation}
where 
\begin{equation}
\mathcal T=\frac{80\pi^{5/2}}{\Gamma\left(\frac{1}{3}\right)^3\Gamma\left(\frac{1}{6}\right)^3d}\eqsim 
\frac{0.422059}{d}
\end{equation}

In anisotropic phase, $\epsilon^2\alpha^2$ correction comes into the black brane mass, $M$ by considering backreactions from the vector order, $\omega(r)$. The black brane mass in anisotropic phase is
\begin{equation}
M^{aniso}=M^{iso}+\frac{281}{1260}\varepsilon^2\alpha^2,
\end{equation}
which can be read off from the metric factor $N(r)$ as we discussed above. By using black brane temperature(\ref{temp-bb}) and (\ref{temp-critical-bb}), we obtain
\begin{equation}
\Delta_\varepsilon E_i\equiv E_i^{iso}-E_i^{aniso}=\frac{843\pi}{34\kappa^2_5}\Sigma_id T_c\left(1-\frac{T}{T_c} \right)^\beta,
\end{equation}
with $\beta=1$.
How much entanglement entropy changes when the vector order, $\omega(r)$ is turned on is given in (\ref{resultyzcritical}). Therefore, the ratio of $\Delta_\varepsilon E_i$ to $\Delta_\varepsilon S_i$ can be computed, which is given by
\begin{equation}
%\frac{\Delta E^{aniso}_i}{\Delta S^{aniso}_i}=
\lim_{d\rightarrow0}\frac{\Delta_\varepsilon E_i}{\Delta_\varepsilon S_i}=\frac{80\pi^{5/2}}{\Gamma\left(\frac{1}{3}\right)^3\Gamma\left(\frac{1}{6}\right)^3d}=\mathcal T,
\end{equation}
where we understand that the entanglement temperature is still universal even in anisotropic phase.

\section{Holographic computation of entanglement entropy of a long cylinder with its radius, $a$ and its length, $L_1$ along $x_1$-direction}
\label{Holographic computation of entanglement entropy of a long cylinder with its radius}
\subsection{Holographic entanglement calculation of cylinder}
\label{Holographic entanglement calculation of cylinder}
%\subsection{The long cylinder along $x_1$-direction}
In this section, we consider a subsystem, of which shape is a {cylinder} lying along the direction of the vector order, $\omega(r)$ on AdS boundary spacetime ($x_1-$direction).\footnote{We note that there are some of entanglement entropy computations on cylinder. Field theory computations of entanglement entropy by employing numerics are in \cite{Huerta:2011qi,Banerjee:2015tia}. For holographic computations, hyperbolic cylinder $R\times H^{d-1}$ is considered\cite{Hung:2011nu} and cylinder in the pure 
AdS background is also considered in \cite{Solodukhin:2008dh}.}
This cylinder is %defined on the AdS boundary, being 
given as, 
$0\leq \rho \leq a$, $0\leq\phi\leq2\pi$ and $-\frac{L_1}{2}\leq x_1 \leq \frac{L_1}{2}$.
The variables, $\rho$ and $\phi$ are $\rho=\sqrt{x^2_2+x^2_3}$ and $\phi=\cot^{-1}\left(\frac{x_2}{x_3}\right)$.  By using these new coordinate variables, we introduce plane polar coordinate on the $x_2-x_3$ plane as
\begin{equation}
ds^2=\frac{1}{z^2}\left\{-z^2N(z)\sigma^2(z)dt^2+\frac{dz^2}{z^2N(z)}+f^{-4}(z)dx^2_1+f^2(z)(d\rho^2+\rho^2d\phi^2)\right\},
\end{equation}
where $z=r^{-1}$, $f(z)$ and $N(z)$ are given in (\ref{fzNzform}), (\ref{Fz_in_fz_form}) and (\ref{N0zNalphazNepsilonzform}).
Now, we consider surface area from the {cylinder} on AdS boundary, where $\phi$ and $x_1$ are to be symmetric directions and we take $L_1$ is to be very large. The surface area is given by
\begin{equation}
\label{area-1-cyl-area}
A^{cy}_1=2\pi L_1\lim_{\delta\rightarrow0}\int^{z^{cy}_*}_{\delta}dz\ z^{-3}\rho(z)\sqrt{\frac{1}{z^2f^2(z)N(z)}+\left(\frac{d\rho}{dz}\right)^2},
\end{equation}
where $z^{cy}_*$ is the maximum value of the coordinate $z$.
To minimize the surface, we apply variation principle to the surface. The condition for the extremization is given by
\begin{equation}
\label{equation-x-cylinder1}
\sqrt{\frac{1}{z^2 f^2(z)N(z)}+(\partial_z \rho)^2}=z^3\frac{\partial}{\partial z}\left(\frac{ \rho(z)\partial_z \rho}{z^3\sqrt{\frac{1}{z^2 f^2(z)N(z)}+(\partial_z \rho)^2}}\right).
\end{equation}

%\footnote
%{$\xi\equiv r^{-2}$;
%\begin{equation}
%A^{cy}_x=2\pi L_x\int_{\varepsilon^2}d\xi\ \frac{\rho(\xi)}{2\xi^2}\sqrt{\frac{1}{\xi f^2(\xi)N(\xi)}+4\xi\left(\frac{d\rho}{d\xi}\right)^2}
%\end{equation}
%Equation of motion:
%\begin{equation}
%\sqrt{\frac{1}{\xi f^2(\xi)N(\xi)}+4\xi\rho'^2(\xi)}=\xi^2\frac{d}{d\xi}\left(\frac{4\rho'(\xi)\rho(\xi)}{\sqrt{\frac{1}{\xi f^2(\xi)N(\xi)}+4\xi\rho'^2(\xi)}}\right)
%\end{equation}
%}

\paragraph{Divergent pieces in the solution, $\rho(z)$}To solve the equation(\ref{equation-x-cylinder1}), we use a fact that the functions $f(z)$ and $N(z)$ have the form of (\ref{fzNzform})% and (\ref{metric-N-factor2}) 
together with an expansion with small $\varepsilon$ and $\alpha$ such that
\begin{equation}
\nonumber
\rho(r)=\rho_0(z)+\alpha^2\rho_\alpha(z)+\varepsilon^2 \alpha^2\rho_\varepsilon(z),
\end{equation}
up to leading order in $\varepsilon^2 \alpha^2$. The solution can be obtained with a form of series solution in small $z$, meaning that an expansion near  boundary, $z=0$. The solutions
% of $\rho_0$ is 
are given by
\begin{equation}
\label{equation-rho-solution-in-r}
\rho_0=a-\frac{z^2}{4a}+O\left(z^4\right),\ 
\rho_\alpha=O\left(z^4\right),\text{\ and\ }
\rho_\epsilon=O\left(z^4\right).
\end{equation}

This solution specifies the divergent pieces of the minimized surface area, which is given by $A_1^{cy}\sim2\pi L_1\lim_{\delta\rightarrow0}\left(\frac{a}{2\delta ^2}+\frac{1}{8a}\log\delta\right)+$finite, where the $\delta$ is again the radial cutoff near AdS boundary \cite{Solodukhin:2008dh}.
% such that $r\rightarrow\frac{1}{\delta}$ as $r$ approaches the boundary. 
%We check such properties later.% will come out once we apply the new expansion of the solution.
%(\ref{equation-cylinder-z-solution}).

{\it{{In our computation, our purpose is to obtain finite pieces of the entanglement entropy of the cylinder (its minimal area).}}} To do this, we need to impose correct boundary conditions at $z=z^{cy}_*$
% (the maximum depth of the stretched surface to deep in the bulk) 
as well as $z={\delta}$ (on the AdS boundary). However, the solution, (\ref{equation-rho-solution-in-r}) will require a boundary condition at $z=z^{cy}_*$ as
\begin{equation}
\nonumber
\left.\frac{\partial\rho}{\partial z}\right\rvert^{z=z^{cy}_*}=\infty,
\end{equation}
since the variable $z$ has a maximum at $z=z^{cy}_*$. This boundary condition is impossible to impose. 

\paragraph{The boundary conditions for the inverse solution, $z(\rho)$}Therefore, we may try an inverse solution, $z(\rho)$.
%note 20220705 43ver.
To get the solution, we define $z(\rho)$ as a series expansion in $\rho$,
\begin{equation}
    z(\rho) \equiv  z^{cy}_*+\sum^{\infty}_{n=1}b_n \rho^n
\end{equation}
together with boundary conditions
\begin{align}
\label{cylinder-z-solution-bc1}
1.&\ z(\rho=0)=z^{cy}_* \quad \text{(automatically satisfied)} \\
\label{cylinder-z-solution-bc2}
2.&\ z(\rho=a)=z^{cy}_*+\sum^{\infty}_{n=1}b_n a^n=0,
\end{align}
where $b_n$ are coefficients of the expansion and $z^{cy}_*$ is the maximum value of the coordinate $z$. %depth of the stretched area deep in the bulk.

To find $z^{cy}_*$, we need to solve a large degree polynomial equation(\ref{cylinder-z-solution-bc2}) (practically n-degree by truncation), which is given by
\begin{equation}
    z^{cy}_*=-\sum^{\infty}_{n=1}a^nb_n.
\end{equation}
The final boundary condition at the turning point($z=z^{cy}_*$) is $\left.\frac{dz(\rho)}{d\rho}\right\rvert^{\rho=0}=0$ being given by
\begin{equation}
\label{cylinder-z-solution-bc3}
    3.\ \left. \frac{dz(\rho)}{d\rho} \right\rvert^{\rho=0}=\left.  \sum^{\infty}_{n=1}nb_n\rho^{n-1} \right\rvert^{\rho=0}=0,
\end{equation}
and that requires $b_1=0$. Therefore, the form of the solution is
\begin{equation}
    z(\rho)=z^{cy}_*+ \sum^{\infty}_{n=2}b_n \rho^n.
\end{equation}
%%%%%%%%%%%%%%%%%%%%%%%upto this 2022.10.07%%%%%%%%%%%%%%5
The solution, $z(\rho)$ will have the following terms in it:
\begin{equation}
\label{equation-cylinder-z-solution}
z(\rho)=z_0(\rho)+\alpha^2z^{(\alpha)} (\rho)+\varepsilon^2\alpha^2z^{(\varepsilon)}(\rho),
\end{equation}
where $z_0$ is the solution in the background of black brane, and $z^{(\alpha)}$ is $\alpha^2$ correction and finally $z^{(\varepsilon)}$ is the $\varepsilon^2\alpha^2$ correction near the critical point, $T=T_c$. Each $z_0(\rho)$, $z^{(\alpha)}(\rho)$, and $z^{(\varepsilon)}(\rho)$ is a series solution in $\rho$ and so satisfies the same boundary conditions (\ref{cylinder-z-solution-bc1}), (\ref{cylinder-z-solution-bc2}) and (\ref{cylinder-z-solution-bc3}) as we have discussed above. 
%$z(\rho)=r^{-1}(\rho)$.
We note that we define the maximum value of each solution as $z_0(0)=z_*$,  $z^{(\alpha)}(0)=z^\alpha_*$ and $z^{(\varepsilon)}(0)=z^\varepsilon_*$ and so
\begin{equation}
\label{equation-cylinder-z-solution-the maximum}
{z^{cy}_*}=z_*+\alpha^2z_*^\alpha+\varepsilon^2\alpha^2z_*^\varepsilon.
\end{equation}

\paragraph{Getting solutions}Now, we illustrate the procedure how to get the solutions.
\begin{itemize}
\item We substitute the trial solution(\ref{equation-cylinder-z-solution}) into equation(\ref{equation-x-cylinder1}) and get series solutions for $z_0$, $z^{(\alpha)}$, and  $z^{(\varepsilon)}$, which are given in Appendix \ref{appendixlabel-B1} in detail. We obtain each series solution upto $O(\rho ^{30})$. %With these solutions, we illustrate system's properties.\\
First, by applying the boundary condition (\ref{cylinder-z-solution-bc2}) to $z_0(\rho)$ solution, we get the ratio of $a$ to $z_*$, i.e. $\frac{a}{z_*}$, with a given $z_*$. By using this information, we can get the value of $a$ with the given $z_*$. We summarize some of the results in Table \ref{Tabble-1}.
\item We also solve equations for $z^{(\alpha)}(\rho)$ and $z^{(\varepsilon)}(\rho)$. The equations and the solutions are given in Appendix \ref{appendixlabel-B1}.
With the given values of $a$ that we obtained from $z_0(\rho)$ solution, 
%Secondly, by keeping these ratios (meaning that keeping the cylinder's radius $"a"$), 
we compute how much the turning point, $z_*^{cy}$(maximum value of $z$) changes by $\alpha ^2$ and $\varepsilon ^2\alpha ^2$ corrections by employing the solutions of $z^{(\alpha)}(\rho)$ and $z^{(\varepsilon)}(\rho)$. These values are also given in Table \ref{Tabble-1}.
\end{itemize}

\setlength\heavyrulewidth{0.25ex}
\newcolumntype{?}{!{\vrule width 2pt}}

\begin{table}[]
\resizebox{\columnwidth}{!}{%
\begin{tabular}{|c|l|l|l|l|l|l|}
\noalign{\hrule height 2pt}
	\multicolumn{1}{?c?}{$\boldsymbol{z_*}$} &
	\multicolumn{1}{c|}{\textbf{0}} &
	\multicolumn{1}{c|}{\textbf{0.05}} &
	\multicolumn{1}{c|}{\textbf{0.10}} &
	\multicolumn{1}{c|}{\textbf{0.15}} &
	\multicolumn{1}{c|}{\textbf{0.20}} &
	\multicolumn{1}{c?}{\textbf{0.25}} \\
\noalign{\hrule height 2pt}
	\multicolumn{1}{?c?}{$\boldsymbol{a}$} &
	\multicolumn{1}{r|}{0} &
	\multicolumn{1}{r|}{0.0398102} &
	\multicolumn{1}{r|}{0.0796221} &
	\multicolumn{1}{r|}{0.119444} &
	\multicolumn{1}{r|}{0.159297} &
	\multicolumn{1}{r?}{0.199223} \\
\hline
	\multicolumn{1}{?c?}{\begin{tabular}[c]{@{}c@{}}$\boldsymbol{a/z_*}$\\ \textbf{deg:38}
							\end{tabular}} &
	\multicolumn{1}{r|}{0.796204} &
	\multicolumn{1}{r|}{0.796205} &
	\multicolumn{1}{r|}{0.796221} &
	\multicolumn{1}{r|}{0.796293} &
	\multicolumn{1}{r|}{0.796485} &
	\multicolumn{1}{r?}{0.796891} \\
\hline
	\multicolumn{1}{?c?}{\begin{tabular}[c]{@{}c@{}}$\boldsymbol{z_*^\alpha}$\\ \textbf{deg:30}
							\end{tabular}} &
	\multicolumn{1}{r|}{0} &
	\multicolumn{1}{r|}{-7.34078*$10^{-7}$} &
	\multicolumn{1}{r|}{-2.33482*$10^{-5}$} &
	\multicolumn{1}{r|}{-1.75502*$10^{-4}$} &
	\multicolumn{1}{r|}{-7.28959*$10^{-4}$} &
	\multicolumn{1}{r?}{-2.18304*$10^{-3}$} \\
\hline 
	\multicolumn{1}{?c?}{\begin{tabular}[c]{@{}c@{}}$\boldsymbol{z_*^\varepsilon}$\\
							\textbf{deg:30}\end{tabular}} &
	\multicolumn{1}{r|}{0} &
	\multicolumn{1}{r|}{-1.52560*$10^{-8}$} &
	\multicolumn{1}{r|}{-4.76626*$10^{-7}$} &
	\multicolumn{1}{r|}{-3.47876*$10^{-6}$} &
	\multicolumn{1}{r|}{-1.38778*$10^{-5}$} &
	\multicolumn{1}{r?}{-3.95175*$10^{-5}$} \\
\noalign{\hrule height 2pt}
	\multicolumn{1}{?c?}{$\boldsymbol{z_*}$} &
	\multicolumn{1}{c|}{\textbf{0.30}} &
	\multicolumn{1}{c|}{\textbf{0.35}} &
	\multicolumn{1}{c|}{\textbf{0.40}} &
	\multicolumn{1}{c|}{\textbf{0.45}} &
	\multicolumn{1}{c|}{\textbf{0.50}} &
	\multicolumn{1}{c?}{} \\
\noalign{\hrule height 2pt}
	\multicolumn{1}{?c?}{$\boldsymbol{a}$} &
	\multicolumn{1}{r|}{0.239290} &
	\multicolumn{1}{r|}{0.279601} &
	\multicolumn{1}{r|}{0.320305} &
	\multicolumn{1}{r|}{0.361608} &
	\multicolumn{1}{r|}{0.403790} &
	\multicolumn{1}{r?}{} \\
\hline
	\multicolumn{1}{?c?}{\begin{tabular}[c]{@{}c@{}}$\boldsymbol{a/z_*}$\\ \textbf{deg:38}
							\end{tabular}} &
	\multicolumn{1}{r|}{0.797632} &
	\multicolumn{1}{r|}{0.798860} &
	\multicolumn{1}{r|}{0.800763} &
	\multicolumn{1}{r|}{0.803573} &
	\multicolumn{1}{r|}{0.807580} &
	\multicolumn{1}{r?}{} \\
\hline
	\multicolumn{1}{?c?}{\begin{tabular}[c]{@{}c@{}}$\boldsymbol{z_*^\alpha}$\\ \textbf{deg:30}
							\end{tabular}} &
	\multicolumn{1}{r|}{-5.30576*$10^{-3}$} &
	\multicolumn{1}{r|}{-1.11452*$10^{-2}$} &
	\multicolumn{1}{r|}{-2.10030*$10^{-2}$} &
	\multicolumn{1}{r|}{-3.63615*$10^{-2}$} &
	\multicolumn{1}{r|}{-5.87538*$10^{-2}$} &
	\multicolumn{1}{c?}{} \\
\hline
	\multicolumn{1}{?c?}{\begin{tabular}[c]{@{}c@{}}$\boldsymbol{z_*^\varepsilon}$\\
							\textbf{deg:30}\end{tabular}} &
	\multicolumn{1}{r|}{-9.05120*$10^{-5}$} &
	\multicolumn{1}{r|}{-1.77833*$10^{-4}$} &
	\multicolumn{1}{r|}{-3.11644*$10^{-4}$} &
	\multicolumn{1}{r|}{-4.99873*$10^{-4}$} &
	\multicolumn{1}{r|}{-7.47390*$10^{-4}$} &
	\multicolumn{1}{r?}{} \\
\noalign{\hrule height 2pt}
\end{tabular}%
}
\caption{In this table, we list the values of $a$(the radius of the cylinder), $z_*^\alpha$(the $\alpha^2$ correction to the maximum value of $z_*$), and $z_*^\varepsilon$(the $\varepsilon^2\alpha^2$ correction to the maximum value of $z_*$) with given values of $z_*$. 
These values are obtained by solving boundary condition(\ref{cylinder-z-solution-bc2}). Solutions are obtained upto $O(a^{38})$ for $z_0$ solution and upto $O(a^{30})$ for $z^{(\alpha)}$ and $z^{(\varepsilon)}$ solutions.
%2022.08.04, new table 1 of 2, solutions of the corrected equation, $z_*^\alpha$ and $z_*^\varepsilon$  (not fitted) }
}
\label{Tabble-1}
\end{table}

\subsection{Evaluation of surface area and subtraction of UV-divergence}
In this subsection, we plug the solutions(\ref{equation-cylinder-z-solution}) into the surface area(\ref{area-1-cyl-area}) to evaluate it(The detailed forms of the solutions are given in (\ref{cylinder-zzero}), (\ref{cylinder-zalpha}) and (\ref{cylinder-zepsilon})).
We notice that once one expands the minimal area by plugging the solution (\ref{equation-cylinder-z-solution}), the divergent pieces of the area will appear. These need to be subtracted. As we did for the case of slabs, we subtract the pure AdS parts from the surface area as follows.
% given in $A_1^{cy(AdS)}$ only, once we expand $A_1^{cy}$ as

Once we evaluate the surface area by employing small $\alpha$ and $\varepsilon$ expansion upto leading order $\alpha^2$ and $\varepsilon^2\alpha^2$, then we have 
\begin{equation}
\nonumber
A_1^{cy}\equiv 2\pi L_1 \left( \gamma^{(0)}+\alpha^2\gamma^{(\alpha)}+\varepsilon^2\alpha^2\gamma^{(\varepsilon)} \right),
\end{equation}
where the divergence pieces are given in $\gamma^{(0)}$ only. When $\alpha=0$ and $\varepsilon=0$, $A_1^{cy}$ becomes $2\pi L_1 \gamma^{(0)}$. $\gamma^{(0)}$ is the surface area in the background of black brane geometry, when $f(z)=\sigma(z)=1$ and $z^2N(z)=1-z^4$. In the factor of $z^2N(z)=1-z^4$, the second terms in $z^2N(z)$ is the effect of black brane mass. Therefore once we consider a new factor,
\begin{equation}
z^2N(z,\xi)\equiv1-\xi z^4.
\end{equation}
We consider such a new metric factor $z^2N(z,\xi)$ and get $\gamma^{(0)}$ again.
Then if $\xi=0$, the solution is in the background of pure AdS, and if $\xi=1$ it becomes in the background of black brane.

To regularize the surface area $\gamma^{(0)}$, we get solutions of $z_0$ again by employing power expansion order by order in $\xi$. This means that we try the following form of the solutions,
\begin{equation}
z_0=z_0^A+\xi z_0^B+O(\xi^2),
\end{equation}
together with the new metric factor, $N(z,\xi)$. The analytic forms of the solutions for $z_0^A$ and $z_0^B$ are given in Appendix \ref{appendixlabel-B2}. Then, we get the surface area $\gamma^{(0)}$ in power expansion in $\xi$, as a form of
\begin{equation}
\gamma^{(0)}=\gamma^{(AdS)}+\xi \gamma^{(\xi)}+O(\xi^2),
\end{equation}
where $\gamma^{(AdS)}$ contains the divergence pieces, which needs to be subtracted. We take the first order in $\xi$ only to estimate the regularized part, and this becomes more accurate near boundary calculation since $z\ll1$ there. Finally we take $\xi=1$. Then, we define $A_1^{cy(ren)}$ as
\begin{equation}
A_1^{cy(ren)}\equiv A_1^{cy}-A_1^{cy(AdS)},
\end{equation}
where $A_1^{cy(AdS)}\equiv 2\pi L_1 \gamma^{(AdS)}$.
Therefore,
%%%%%%%%%%%%%%%%%%2022.10.14%%%%%%%%%%%%%%%%
\begin{equation}
\nonumber
A_1^{cy(ren)}\equiv 2\pi L_1 \left( \gamma^{(\xi)}+\alpha^2\gamma^{(\alpha)}+\varepsilon^2\alpha^2\gamma^{(\varepsilon)} \right),
\end{equation}
and our renormalized entanglement entropy $S^{cy(ren)}$ for the cylinder is given by
\begin{equation}
S_{cy}^{(ren)}=\frac{2\pi}{\kappa^2_5}A_1^{cy(ren)}.
\end{equation}
\paragraph{Entanglement entropy of Cylinder}
$\gamma^{(\xi)}$, $\gamma^{(\alpha)}$ and $\gamma^{(\varepsilon)}$ are graphically obtained in Figure.\ref{fig1}, \ref{fig2} and \ref{fig3} in order. The analytic forms of these are given in Appendix \ref{appendixlabel-B3}. Their leading behaviors when $a$ is small are given by
\begin{eqnarray}
\gamma^{(\xi)}(a)=0.73228506a^3+O(a^4), \\
\gamma^{(\alpha)}(a)=7.8111218a^3+O(a^4), \\
\gamma^{(\varepsilon)}(a)=0.1633128a^3+O(a^4).
\end{eqnarray}
%For the subsystem with its shape of cylinder, we also compute the same quantity, 
Once we define $\Delta_\varepsilon S_{cy}=S^{iso}_{cy}-S^{aniso}_{cy}$ as we discussed in the slab case, we find that 
\begin{equation}
\Delta_\varepsilon S_{cy}=\frac{5040\pi^2L_1}{17\kappa^2_5}\gamma^{(\varepsilon)}(a)T_c\left(1-\frac{T}{T_c}\right)^\beta,
\end{equation}
%where 
where $a$ and $L_1$ is the radius and the length of the cylinder.
Our analytic computation shows that holographic entanglement entropy excess $\Delta_\varepsilon S_{cy}$ for cylinder also presents scaling behavior $\sim\left(1-\frac{T}{T_c}\right)^\beta$ and its critical exponent $\beta =1$.
%We note that to get this results, we utilize analytic as well as numerical methods. 

\paragraph{The first law of entanglement entropy and entanglement temperature}
In figure.\ref{fig4}, we plot the following quantities in order:
\begin{align}
\label{plotted_function_xi}
\frac{1}{a}\frac{\Delta_0 S_{cy}}{\Delta_0 E} &=\frac{8\pi \gamma^{(\xi)}(a)}{3a^3} {\rm \ \ (\xi -graph)}, \\
\label{plotted_function_alpha}
\frac{1}{a}\frac{\Delta_\alpha S_{cy}}{\Delta_\alpha E} &=\frac{\pi \gamma^{(\alpha)}(a)}{4a^3} {\rm \ \ (\alpha^2 -graph)}, \\
\label{plotted_function_epsilon}
\frac{1}{a}\frac{\Delta_\varepsilon S_{cy}}{\Delta_\varepsilon E} &=\frac{3360\pi \gamma^{(\varepsilon)}(a)}{281a^3} {\rm \ \ (\varepsilon^2\alpha^2 -graph)},
\end{align}
and it turns out that as $a$ approach zero, they meet at one point. We again note that $\Delta_0 Q$ represents the difference between the quantity, $Q$ computed in black brane background and in pure AdS background.  $\Delta_\alpha Q$ is the $\alpha^2$ correction to the quantity, $Q$ when the chemical potential is turned on. Finally, $\Delta_\varepsilon$ is $\varepsilon^2\alpha^2$ correction to the quantity, $Q$ when the vector order appears near the critical point. Therefore, we conclude that when $a\rightarrow0$,
\begin{equation}
\lim_{a\rightarrow0}\frac{1}{a}\frac{\Delta S_{cy}}{\Delta E}=\lim_{a\rightarrow0}\frac{1}{a}\frac{\Delta_0 S_{cy}}{\Delta_0 E}=\lim_{a\rightarrow0}\frac{1}{a}\frac{\Delta_\alpha S_{cy}}{\Delta_\alpha E}=\lim_{a\rightarrow0}\frac{1}{a}\frac{\Delta_\varepsilon S_{cy}}{\Delta_\varepsilon E}=c^{-1}_{ent},
\end{equation}
where
\begin{equation}
c_{ent}=0.163004\pm0.000001.
\end{equation} 
This value is obtained by computing the average and standard deviation of the values of the functions (\ref{plotted_function_xi}), (\ref{plotted_function_alpha}), and (\ref{plotted_function_epsilon}) at $a=0$.
By using the definition of entanglement temperature (\ref{definition_of_entanglement_entropy_temperature}), we understand that even in the case that the vector order appears in anisotropic phase, the first law of entanglement entropy is retained.
%where  $a$ is the radius of the cylinder. 
The entanglement temperature is given by
\begin{equation}
\mathcal T_{cy}=\frac{c_{ent}}{a}.
\end{equation}

\begin{figure}[!htb]
\minipage{0.485\textwidth}%
  \includegraphics[width=\linewidth]{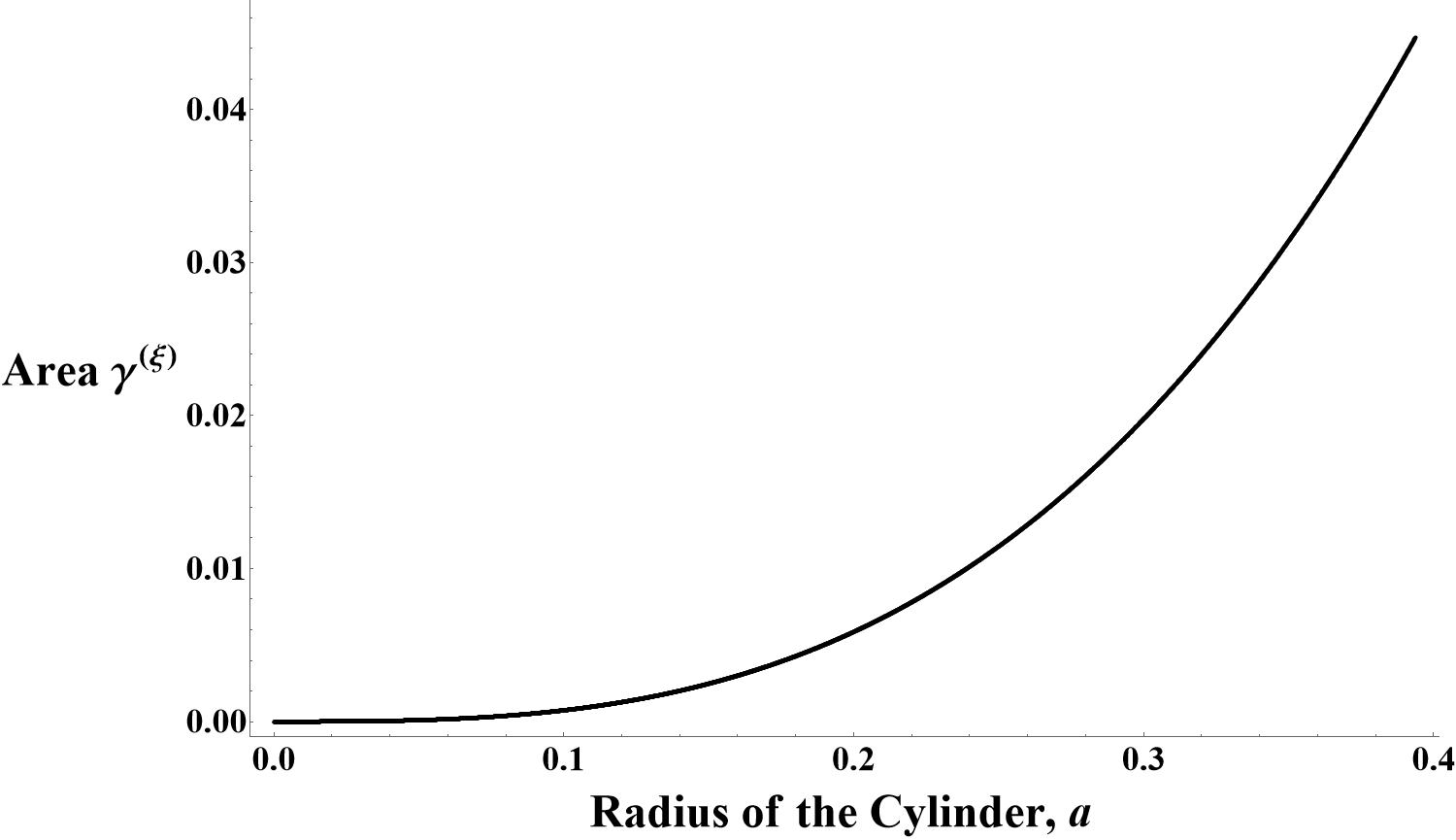}
  \caption{$\gamma^{(\xi)}(a)=0.73228506a^3+\ldots$,\\from $a=7.87386\times 10^{-5}$ to $a=0.393693$,\\5000 data points.}
\label{fig1}
\endminipage
\hfill
\minipage{0.485\textwidth}
  \includegraphics[width=\linewidth]{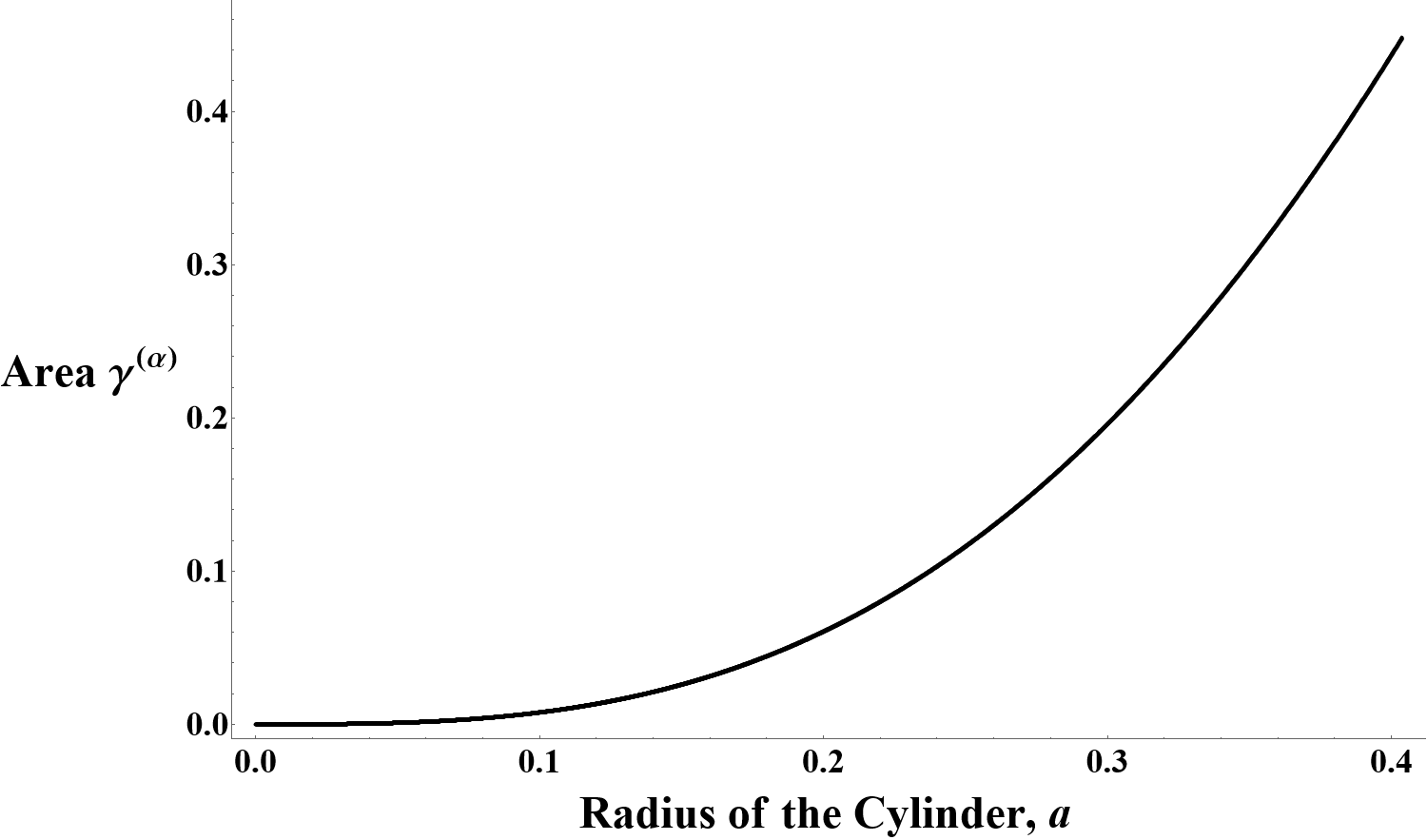}
  \caption{$\gamma^{(\alpha)} (a)=7.8111218a^3+\ldots$,\\from $a=7.96204\times 10^{-5}$ to $a=0.403790$,\\5000 data points.}
\label{fig2}
\endminipage\hfill
\end{figure}
\begin{figure}[!htb]
\minipage{0.485\textwidth}
  \includegraphics[width=\linewidth]{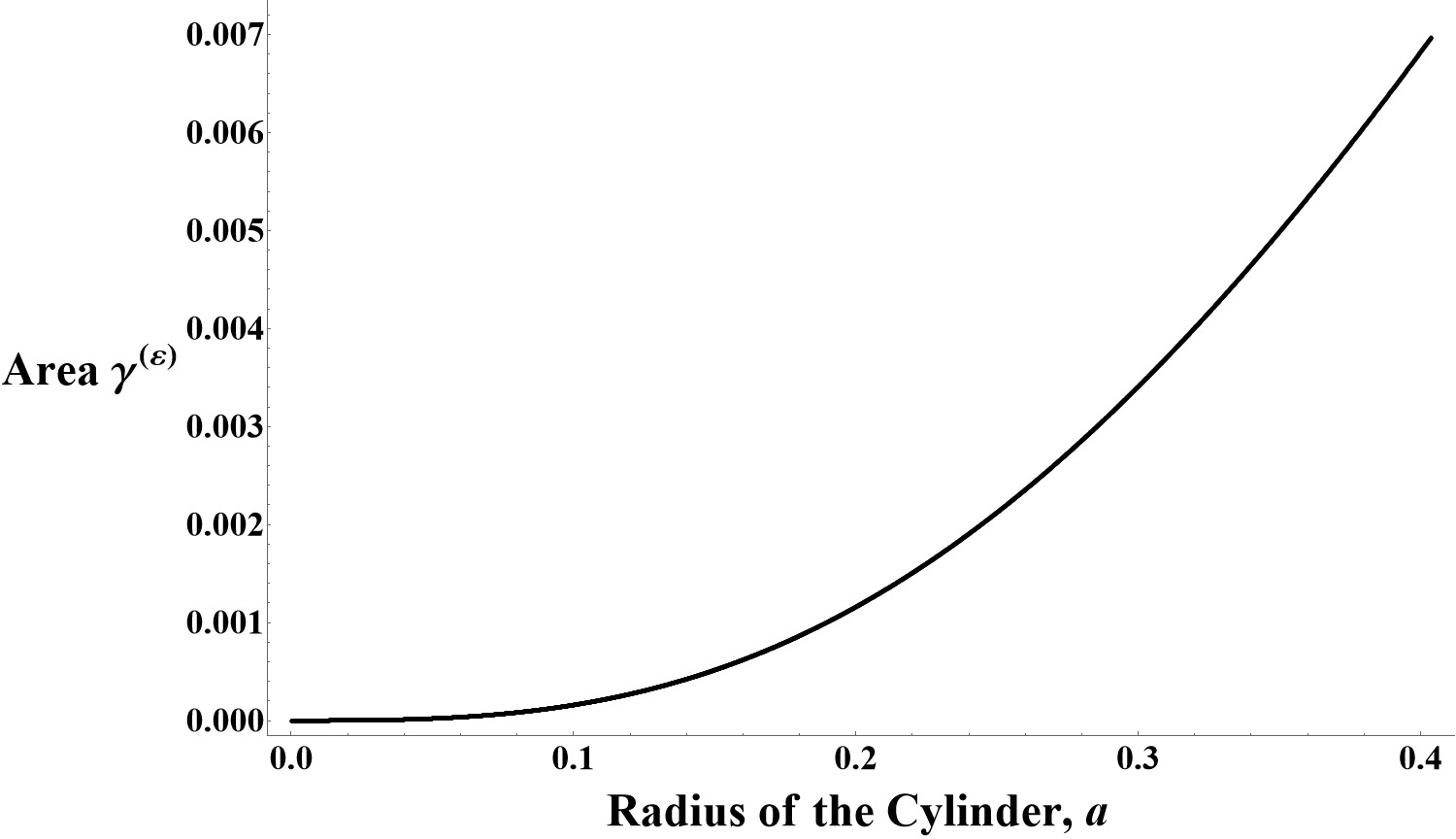}
  \caption{$\gamma^{(\varepsilon)} (a)=0.1633128a^3+\ldots$,\\from $a=7.96204\times 10^{-5}$ to $a=0.403790$,\\5000 data points.}
\label{fig3}
\endminipage
\hfill
\minipage{0.486\textwidth}
  \includegraphics[width=\linewidth]{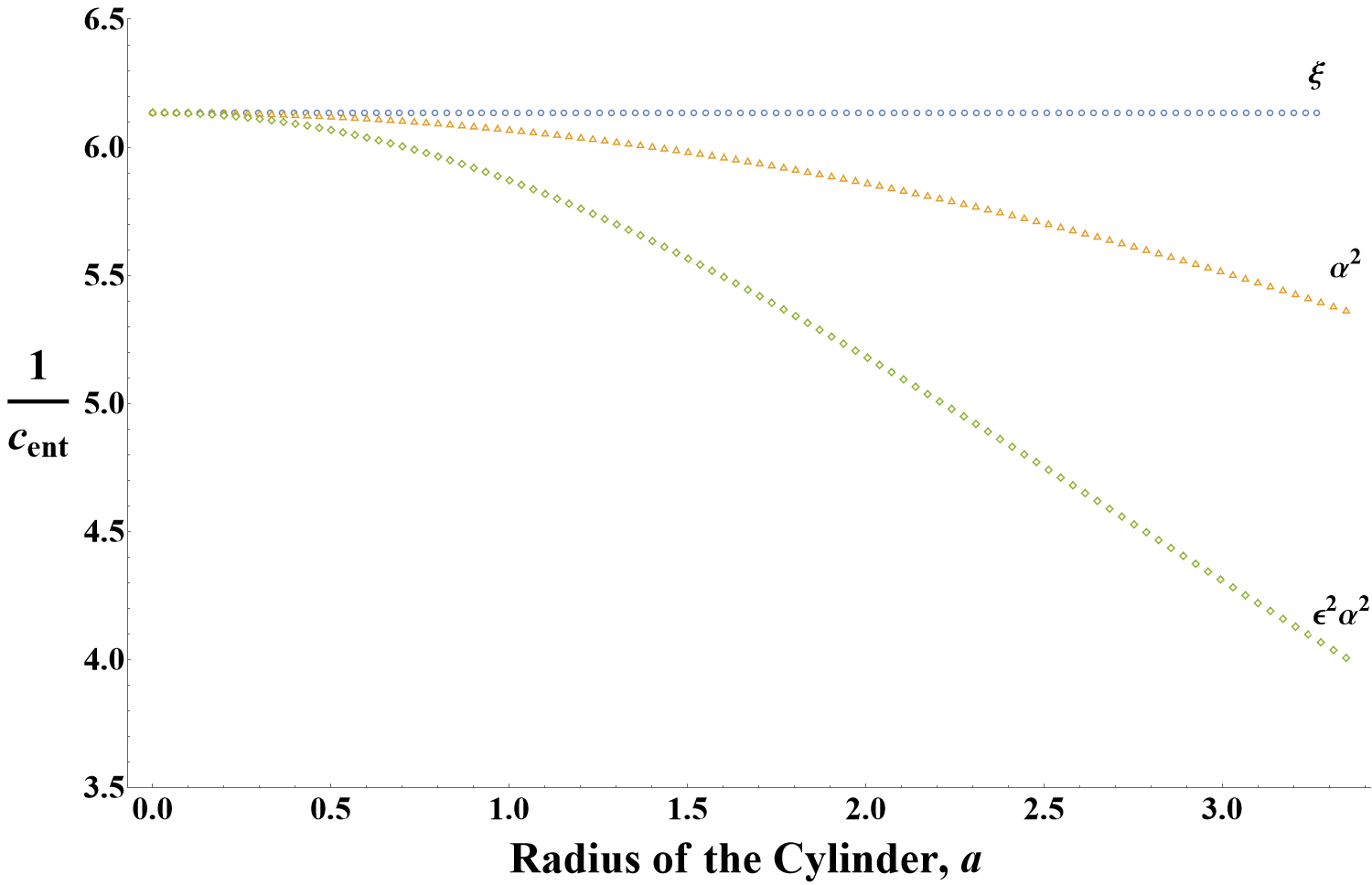}
  \captionsetup{font=footnotesize}
  \caption{$\xi\rightarrow\lim_{a\rightarrow 0} \frac{8\pi \gamma^{(\xi)}(a)}{3a^3}=6.134777049077025$,\\
  $\alpha^2 \rightarrow \lim_{a\rightarrow 0} \frac{\pi \gamma^{(\alpha)}(a)}{4a^3} =6.134840767351422$,\\
  $\varepsilon^2\alpha^2\rightarrow \lim_{a\rightarrow 0} \frac{3360\pi \gamma^{(\varepsilon)}(a)}{281a^3}=6.134840676290574$,\\100 data points each}
\label{fig4}
\endminipage\hfill
\end{figure}

\section{Discussion}
{%\color{blue}
In this paper, we explore 4-dimensional holographic anisotropic super fluids defined on the boundary of 5-dimensional asymptotically AdS spacetime  near its critical point $T=T_c$, where a vector order parameter appears, and it breaks SO(3)-rotational symmetry of spacetime down to SO(2). The gravity dual of such a system is Einstein-SU(2)Yang-Mills theory, defined in asymptotically AdS spacetime. To understand properties of this system, we compute holographic entanglement entropy in the background of the charged black brane solution of this gravity system. We apply an analytic method and obtain holographic entanglement entropies of subsystems with shapes of wide and thin slabs and a long cylinder. 

For the wide and thin slabs, we consider two different spatial directions: one is lying along a direction which is parallel to the vector order whereas another is perpendicular to the vector order. For the cylinder case, we consider the cylinder lying along the vector order only. The entanglement entropies that we obtained for the slab and cylinder cases share universal properties: these show a scaling behavior near critical point, which has a form of
\begin{equation}
\Delta_\varepsilon S\sim\left(1-\frac{T}{T_c}\right)^{\beta},
\end{equation}
where 
$\Delta_\varepsilon S=S^{iso}-S^{aniso}$, and $S^{iso}$ is the entanglement entropy in isotropic phase whereas $S^{aniso}$ is that in anisotropic phase. The critical exponent, $\beta=1$ for all of the cases that we examine. Therefore, we understand that 
the critical exponent $\beta=1$ is probably the common feature of the entanglement entropy near the critical point of the system. We note that the analytic approach is valid when $\alpha=\frac{\kappa_5}{g}$ is small, and in this case, the system undergoes second order phase transition near the critical point. We restrict our analysis in this case only.
%entanglement entropies. %for the slabs and cylinder cases, and $i=1,2$. $i=1$ is for the slab being  parallel to the vector order whereas $i=2$ is for the slab being perpendicular to the vector order.

%(However, more interesting feature of the entanglement entropies is anisotropy, which appears in the coefficients of those. The coefficients mean $C_i(d)$, where for example, we consider the entanglement entropies for the wide and thin slab cases, of which form of those are given by) {\color{red} ($\to$ 
However, an interesting feature occurs when we consider anisotropy. The entanglement entropy of slabs has the following expansion near the critical point
\begin{equation}
\Delta_\varepsilon S_i= C_i(d)\left(1-\frac{T}{T_c}\right)^{\beta},
\end{equation}
where $i=1,2$ and $i=1$ is for the slab being parallel to the vector order whereas $i=2$ is for the slab being perpendicular to the vector order. We assume that the shapes of those two slabs are the {same} but their directions are different. We determine {the coefficient, $C_i(d)$,} by employing small ``$d$'' expansion such as
\begin{equation}
C_i(d)=C^{(2)}_i d^2+ C^{(4)}_i d^4+...
\end{equation} 
where the $d$ is the thickness of the slabs. In the small $d$ region, we probe ultraviolet degrees of freedom and it corresponds to that the extended minimal surface to the bulk from the slab is still probing the bulk region near the AdS boundary. However, condensation is an infrared effect. This means that we need to probe deeper in the infrared {region} to see the effects of condensation. In its holographic dual, as $d$ grows, the extended minimal surface to the bulk from the slab probes deeper in the bulk.

In the small $d$ region, the term being proportional to $C^{(2)}$ is the most dominant and it turns out that $C^{(2)}_1=C^{(2)}_2$.  This implies that the ultraviolet degrees of freedom is still universal in a sense that they 
%{\color{red} (have no information about) $\to$ 
are independent of the directions of the vector order parameter.
% being emergent near critical point. 
An interesting anisotropy appears in $C^{(4)}_i$ such that $C^{(4)}_1\neq C^{(4)}_2$. This observation leads us to define a new order parameter from entanglement entropy, which is defined as
\begin{equation}
\mathcal O_{12}\equiv S_1-S_2.
\end{equation}
We understand that $O_{12}=0$ above the critical temperature since SO(3)-rotational symmetry is retained. However, $O_{12}$ shows critical behavior together with its non-zero coefficient near the critical point and it can be an indication of phase transition.
%implies phase transition since it breaks the SO(3) rotational symmetry.

With the same reason, the condensation does not spoil the first law of entanglement entropy, since the law holds in the $d\rightarrow 0$ limit, which implies that it is ultraviolet physics. In the small size limit of the subsystem($d\rightarrow 0$ limit), it turns out that the ratio of entropy change to total energy change of the subsystem is universal.
% {\color{red}(, which depends only on the shape of the subsystem and spacetime geometry even near the critical point)}.  
The term being proportional to $C^{(4)}$ is relatively infrared effect and it is subleading in small $d$ expansion. %{\color{red}(This term does not affect the ultraviolet physics. )}

In conclusion, in anisotropic holographic superfluid system, we find that the system presents universal properties and anisotropy at the same time. The universal properties are scaling behaviors of the entanglement entropies and all of the subsystems that we study  share their critical exponent $\beta=1$. When one looks at ultraviolet degrees of freedom, the first law of entanglement entropy is held for all the subsystems that we look at. This is also a universal feature in a sense that it does not depend on the direction of the vector order parameter. However, if one averts one's eyes to the infrared region, one can see a fact that entanglement entropy depends on the direction of the order parameter. By using this fact, one can define an interesting order parameter near the critical point.

%related to the shape of the subsystem on the AdS boundary and bulk geometry only and this means that  However, the subleading corrections are different, which show directional dependency of the entanglement entropies. 
%
%Condensation is infrared effect. This means that once we consider holographic dual of the system, we need toecomes more significant 
%
%
%
%The directional dependency of the entanglement entrop appears in the subleading corrections since they might be an {\it infrared} effects.
%
}

%\newpage
\section*{Acknowledgement}
J.H.O thanks his W.J. and Y.J.
This work was supported by the National Research Foundation of Korea(NRF) grant funded by the Korea government(MSIT). (No.2021R1F1A1047930). {CP was supported by the National Research Foundation of Korea(NRF) grant funded by the Korea government(MSIT) (No. NRF-2019R1A2C1006639).}

\begin{appendices}
\section{Slabs on the $x_2$-$x_3$ and $x_3$-$x_1$ planes}
%In this appendix we show how the calculations are done, to get the relationship between the minimized area of a slab on a plane and the thickness of the slab, $d$ (the coordinate size in the direction normal to the plane).
For the computations of the slabs, we need to evaluate the following form of the integration:
\begin{equation}
I_n \equiv \int^{\infty}_{1} \frac{u^n}{\sqrt{u^6-1}}\ du,
\end{equation}
where $n$ is an integer.
For some specific $n$, there are some results:
\begin{align}
\nonumber
I_{-8}=\frac{4\sqrt{\pi}\Gamma\left(\frac{2}{3}\right)}{7\ \Gamma\left(\frac{1}{6}\right)},\
I_{-6}=\frac{\sqrt{\pi}\Gamma\left(\frac{1}{3}\right)}{15\ \Gamma\left(\frac{5}{6}\right)},\
I_{-4}=\frac{1}{3},\ 
I_{-2}=\frac{\sqrt{\pi}\ \Gamma\left(\frac{2}{3}\right)}{\Gamma\left(\frac{1}{6}\right)},\ 
I_{0}=\frac{\sqrt{\pi}\ \Gamma\left(\frac{1}{3}\right)}{6\ \Gamma\left(\frac{5}{6}\right)},\\ \nonumber 
I_4=\lim_{\delta\rightarrow 0}\left[\frac{1}{2\delta^2}+O(\delta^3)\right]-\frac{\sqrt{\pi}\ \Gamma\left(\frac{2}{3}\right)}{2\ \Gamma\left(\frac{1}{6}\right)}
\end{align}

We define variables $s=1/r_{*}$ and $u=rs$. In fact, to get the area of the slabs analytically, we expand the metric factors appearing in the calculation in terms of $s$, where we expand $f(r)$ and $N(r)$ in terms of $s$. They are given by
\begin{align}
f&=1-\frac{1}{18}\frac{(1-2 r^2)}{\left(1+r^2\right)^4}\ \varepsilon^2\alpha^2\\ \nonumber
&=1-\frac{1}{18}\varepsilon^2\alpha^2\left(1-\frac{2u^2}{s^2}\right)\left(1+\frac{u^2}{s^2}\right)^{-4}\\ \nonumber
&=1+\frac{\varepsilon^2\alpha^2}{9 u^6}s^6-\frac{\varepsilon^2\alpha^2}{2u^8}s^8+O(s^9)
\end{align}
and
\begin{align}
N&=r^2-\frac{1}{r^2}+\frac{32}{3} \alpha^2\left(\frac{1}{r^4}-\frac{1}{r^2}\right)-\frac{4\varepsilon^2\alpha^2}{9r^2}\left[\frac{281}{560}\left(1-\frac{1}{r^2}\right)-\frac{3r^2}{2(1+r^2)^2}+\frac{1+2r^2}{r^2(1+r^2)^3}\right]\\ \nonumber
&=r^2\left(1-\frac{1}{r^4}\right)\left[1-\frac{32}{3}\alpha^2r^{-2}(1+r^2)^{-1}+\frac{\varepsilon^2\alpha^2}{1260}(279r^{-2}+837-3r^2-281r^4)(1+r^2)^{-4}\right]\\ \nonumber
&=\frac{u^2}{s^2}\left[1-\left(1+\frac{32}{3}\alpha^2+\frac{281}{1260}\varepsilon^2\alpha^2\right)\frac{s^4}{u^4}+\left(\frac{32}{3}\alpha^2+\frac{1121}{1260}\varepsilon^2\alpha^2\right)\frac{s^6}{u^6}-\frac{4\varepsilon^2\alpha^2}{3}\frac{s^8}{u^8}\right.\\ \nonumber
&\ \ \left.+\frac{10\varepsilon^2\alpha^2}{9}\frac{s^{10}}{u^{10}}\right]+O(s^9)
\end{align}

\subsection{Slab on AdS boundary and divergence subtraction}
%We verify that the area of the slab in the ground state, $A_1^{(AdS)}$, cancels out not only the divergent term of $A_1$, but also the inverse square term. In AdS limit,
For surface area computation in AdS space, we use metric(\ref{background-metruc}) but the metric factors of $f$, $\sigma$, and $N$ are replaced by
\begin{equation}
\nonumber
\sigma =1,\ f=1,\ N(r)=r^2.
\end{equation}
Then, the relations between $d$, $s$, and the surface area $A_1^{(AdS)}$ are given by 
\begin{equation}
\nonumber
d=2\int^\infty_{r_*}\frac{dr}{\sqrt{\frac{r^2N(r)}{f^4}\left(\frac{r^6}{r^6_*}-1\right)}}
=2\int^\infty_{r_*} \frac{dr}{r^2\sqrt{\frac{r^6}{r^6_*}-1}}
=\frac{2\sqrt{\pi}\ \Gamma\left(\frac{2}{3}\right)}{\Gamma\left(\frac{1}{6}\right)}\ s,
\end{equation}
namely,
\begin{equation}
\nonumber
s(d)=\frac{\Gamma\left(\frac{1}{6}\right)}{2\sqrt{\pi}\ \Gamma\left(\frac{2}{3}\right)}\ d.
\end{equation}
The surface area $A_1^{(AdS)}$ is given by
\begin{align}
\nonumber
A_1^{(AdS)}&=2L_2L_3\int^\infty_{r_*}dr \frac{r^2f^2}{\sqrt{N(r)}}\left(1-\frac{r^6_*}{r^6}\right)^{-\frac{1}{2}}
=2\Sigma_1\int^\infty_{r_*}dr \frac{r}{\sqrt{1-\frac{r^6_*}{r^6}}}
\\ \nonumber
&=\lim_{\delta\rightarrow 0}\frac{1}{\delta^2}
-\frac{\sqrt{\pi}\Gamma\left(\frac{2}{3}\right)}{\Gamma\left(\frac{1}{6}\right)}\frac{1}{s^2}
\\ \nonumber
&=\lim_{\delta\rightarrow 0}\frac{1}{\delta^2}+C_{23}^{-2}\frac{1}{d^2}.
\end{align}
%Repeating the same process we obtain
%\begin{align}
%\nonumber
%s(d)&=\frac{\Gamma\left(\frac{1}{6}\right)}{2\sqrt{\pi}\ \Gamma\left(\frac{2}{3}\right)}\ d
%\end{align}
%and
%\begin{align}
%\nonumber
%\frac{A_1^{CFT}}{L_2L_3}&=
%\lim_{\delta\rightarrow 0}\frac{1}{\delta^2}
%-\frac{\sqrt{\pi}\Gamma\left(\frac{2}{3}\right)}{\Gamma\left(\frac{1}{6}\right)}\frac{1}{s^2}
%\\ \nonumber
%&=\lim_{\delta\rightarrow 0}\frac{1}{\delta^2}+C_{23}^{-2}\frac{1}{d^2}.
%\end{align}
Finally, we get
\begin{equation}
\frac{A_1^{(ren)}}{\Sigma_1} \equiv \frac{A_1-A_1^{(AdS)}}{\Sigma_1}=C^2_{23}d^2+C^4_{23}d^4+C^6_{23}d^6+O(d^7).
\end{equation}

\subsection{Slab on the $x_2$-$x_3$ plane}
The relation between the thickness of the slab ``$d$" and $r_*$, maximal depth of the stretched surface (or, turning point) for the slab on the $x_2$-$x_3$ plane is given by
\begin{align}
d&=2\int^\infty_{r_*}\frac{dr}{\sqrt{\frac{r^2N(r)}{f^4(r)}\left(\frac{r^6}{r_*^6}-1\right)}} =\int^\infty_1du\ \frac{2}{u\sqrt{u^6-1}}\frac{f^2(u/s)}{\sqrt{N(u/s)}}\\ \nonumber
&=\int^\infty_1du\ \frac{2}{u\sqrt{u^6-1}}\left(1+\frac{2\varepsilon^2\alpha^2}{9 u^6}s^6-\frac{\varepsilon^2\alpha^2}{u^8}s^8\right)\frac{s}{u}\left[1+\frac{1}{2}\left(1+\frac{32}{3}\alpha^2+\frac{281}{1260}\varepsilon^2\alpha^2\right)\frac{s^4}{u^4}\right.\\ \nonumber 
&-\left.\frac{1}{2}\left(\frac{32}{3}\alpha^2+\frac{1121}{1260}\varepsilon^2\alpha^2\right)\frac{s^6}{u^6}+\frac{2\varepsilon^2\alpha^2}{3}\frac{s^8}{u^8}\right]\\ \nonumber
&=(2I_{-2})\ s+I_{-6}\left(1+\frac{32}{3}\alpha^2+\frac{281}{1260}\varepsilon^2\alpha^2\right)s^5-I_{-8}\left(\frac{32}{3}\alpha^2+\frac{187}{420}\varepsilon^2\alpha^2\right)s^7+O(s^8)\\ \nonumber
&=\frac{2\sqrt{\pi}\ \Gamma\left(\frac{2}{3}\right)}{\Gamma\left(\frac{1}{6}\right)}\ s+\frac{\sqrt{\pi}\ \Gamma\left(\frac{1}{3}\right)}{15\Gamma\left(\frac{5}{6}\right)}\left(1+\frac{32}{3}\alpha^2+\frac{281}{1260}\varepsilon^2\alpha^2\right)s^5-\frac{4\sqrt{\pi}\ \Gamma\left(\frac{2}{3}\right)}{7\Gamma\left(\frac{1}{6}\right)}\left(\frac{32}{3}\alpha^2+\frac{187}{420}\varepsilon^2\alpha^2\right)s^7+O(s^8),
\end{align}
where we expand the integration in terms of $s$ upto its 7th order.\\
From this form of the expansion, the expression for $s$ can be given in series of small $d$.
\begin{align}
\label{x2x3plane_s}
s(d)&=\frac{\Gamma\left(\frac{1}{6}\right)}{2\sqrt{\pi}\ \Gamma\left(\frac{2}{3}\right)}\ d - \frac{5\Gamma\left(\frac{1}{3}\right)\Gamma\left(\frac{1}{6}\right)^6}{6912\sqrt{\pi^5}\ \Gamma\left(\frac{2}{3}\right)^6\Gamma\left(\frac{5}{6}\right)}\left(1+\frac{32}{3}\alpha^2+\frac{281}{1260}\varepsilon^2\alpha^2\right)d^5 \\ \nonumber
&+ \frac{\Gamma\left(\frac{1}{6}\right)^7}{448\sqrt{\pi^7}\Gamma\left(\frac{2}{3}\right)^7}\left(\frac{32}{3}\alpha^2+\frac{187}{420}\varepsilon^2\alpha^2\right)d^7+O(d^8)
\end{align}
$A_1$ in terms of $s$, upto 6th order is given by.
\begin{align}
\label{x2x3plane_area}
\frac{A_1}{L_2L_3}&=2\int^\infty_{r_*} dr\ \frac{r^2f^2(r)}{\sqrt{N(r)}}\left(1-\frac{r^6_*}{r^6}\right)^{-\frac{1}{2}}\\ \nonumber
&=\int^\infty_1du\ \frac{2u^5}{\sqrt{u^6-1}}\frac{1}{s^3}\frac{f^2(u/s)}{\sqrt{N(u/s)}}\\ \nonumber
&= 2I_4\frac{1}{s^2}+I_0\left(1+\frac{32}{3}\alpha^2+\frac{281}{1260}\varepsilon^2\alpha^2\right)s^2 -I_{-2}\left(\frac{32}{3}\alpha^2+\frac{187}{420}\varepsilon^2\alpha^2\right)s^4\\ \nonumber
&+I_{-4}\left(\frac{3}{4}+16\alpha^2-\frac{93}{280}\varepsilon^2\alpha^2\right)s^6+O(s^7)\\ \nonumber
&=\lim_{\delta\rightarrow 0}\frac{1}{\delta^2}-\frac{\sqrt{\pi}\Gamma\left(\frac{2}{3}\right)}{\Gamma\left(\frac{1}{6}\right)}\frac{1}{s^2}+\frac{\sqrt{\pi}\Gamma\left(\frac{1}{3}\right)}{6\ \Gamma\left(\frac{5}{6}\right)}\left(1+\frac{32}{3}\alpha^2+\frac{281}{1260}\varepsilon^2\alpha^2\right)s^2\\ \nonumber
&-\frac{\sqrt{\pi}\Gamma\left(\frac{2}{3}\right)}{\ \Gamma\left(\frac{1}{6}\right)}\left(\frac{32}{3}\alpha^2+\frac{187}{420}\varepsilon^2\alpha^2\right)s^4+\left(\frac{1}{4}+\frac{16}{3}\alpha^2-\frac{31}{280}\varepsilon^2\alpha^2\right)s^6 + O(s^7)
\end{align}
Finally, by using the relations (\ref{x2x3plane_s}) and (\ref{x2x3plane_area}), we get
\begin{align}
\label{appendix-slab-A1}
\frac{A_1}{L_2L_3}&= \lim_{\delta\rightarrow 0}\frac{1}{\delta^2}+C^{-2}_{23}\frac{1}{d^2}+C^2_{23}d^2+C^4_{23}d^4+C^6_{23}d^6+O(d^7)
\end{align}
where
\begin{align}
\nonumber
C^{-2}_{23}&=-\frac{32\sqrt{\pi^9}}{3\sqrt{3}}\frac{1}{\Gamma\left(\frac{1}{3}\right)^3\Gamma\left(\frac{1}{6}\right)^3}\\ \nonumber
C^2_{23}&=\frac{3}{320\sqrt{\pi^7}}\Gamma\left(\frac{1}{3}\right)^3\Gamma\left(\frac{1}{6}\right)^3\left(1+\frac{32}{3}\alpha^2+\frac{281}{1260}\varepsilon^2\alpha^2\right)\\ \nonumber
C^4_{23}&=-\frac{3\sqrt{3}}{28\sqrt{\pi^9}}\Gamma\left(\frac{1}{3}\right)^3\Gamma\left(\frac{1}{6}\right)^3\left(\alpha^2+\frac{187}{4480}\varepsilon^2\alpha^2\right)\\ \nonumber
C^6_{23}&=\frac{27}{16384\pi^9}\Gamma\left(\frac{1}{3}\right)^6\Gamma\left(\frac{1}{6}\right)^6\left(1-\frac{13\sqrt{3}}{1800\pi^{5/2}}\Gamma\left(\frac{1}{3}\right)^3\Gamma\left(\frac{1}{6}\right)^3\right)\\ \nonumber
&+\frac{9}{256\pi^9}\Gamma\left(\frac{1}{3}\right)^6\Gamma\left(\frac{1}{6}\right)^6\left(1-\frac{13\sqrt{3}}{1800\pi^{5/2}}\Gamma\left(\frac{1}{3}\right)^3\Gamma\left(\frac{1}{6}\right)^3\right)\alpha^2\\ \nonumber
&-\frac{837}{1146880\pi^9}\Gamma\left(\frac{1}{3}\right)^6\Gamma\left(\frac{1}{6}\right)^6\left(1+\frac{3653\sqrt{3}}{502200\pi^{5/2}}\Gamma\left(\frac{1}{3}\right)^3\Gamma\left(\frac{1}{6}\right)^3\right)\varepsilon^2\alpha^2,
\end{align}
with $\Gamma\left(\frac{2}{3}\right)$ and $\Gamma\left(\frac{5}{6}\right)$, which are
\begin{align}
\nonumber
\Gamma\left(\frac{2}{3}\right)&=\frac{2\pi}{\sqrt{3}\ \Gamma\left(\frac{1}{3}\right)},\\ \nonumber
\Gamma\left(\frac{5}{6}\right)&=\frac{2\pi}{\Gamma\left(\frac{1}{6}\right)}.
\end{align}

\subsection{Slab on the $x_3$-$x_1$ plane}
In this subsection we find the expression of $A_2$ in terms of $d$, following the same steps as we did in A.1. First, $d$ is expanded in terms of $s$ upto its 7th order. 
\begin{align}
d&=2\int^\infty_{r_*}\frac{dr}{\sqrt{r^2f^2(r)N(r)\left(\frac{r^6}{r_*^6}-1\right)}}
=\int^\infty_1du\ \frac{2}{u\sqrt{u^6-1}}\frac{1}{f(u/s)\sqrt{N(u/s)}}\\ \nonumber
&=\int^\infty_1du\ \frac{2}{u\sqrt{u^6-1}}\left(1-\frac{\varepsilon^2\alpha^2}{9 u^6}s^6+\frac{\varepsilon^2\alpha^2}{2u^8}s^8\right)\frac{s}{u}\left[1+\frac{1}{2}\left(1+\frac{32}{3}\alpha^2+\frac{281}{1260}\varepsilon^2\alpha^2\right)\frac{s^4}{u^4}\right.\\ \nonumber
&-\left.\frac{1}{2}\left(\frac{32}{3}\alpha^2+\frac{1121}{1260}\varepsilon^2\alpha^2\right)\frac{s^6}{u^6}+\frac{2\varepsilon^2\alpha^2}{3}\frac{s^8}{u^8}\right]\\ \nonumber
&=(2I_{-2})\ s+I_{-6}\left(1+\frac{32}{3}\alpha^2+\frac{281}{1260}\varepsilon^2\alpha^2\right)s^5-I_{-8}\left(\frac{32}{3}\alpha^2+\frac{467}{420}\varepsilon^2\alpha^2\right)s^7+O(s^8)\\ \nonumber
&=\frac{2\sqrt{\pi}\ \Gamma\left(\frac{2}{3}\right)}{\Gamma\left(\frac{1}{6}\right)}\ s+\frac{\sqrt{\pi}\ \Gamma\left(\frac{1}{3}\right)}{15\Gamma\left(\frac{5}{6}\right)}\left(1+\frac{32}{3}\alpha^2+\frac{281}{1260}\varepsilon^2\alpha^2\right)s^5-\frac{4\sqrt{\pi}\ \Gamma\left(\frac{2}{3}\right)}{7\Gamma\left(\frac{1}{6}\right)}\left(\frac{32}{3}\alpha^2+\frac{467}{420}\varepsilon^2\alpha^2\right)s^7+O(s^8)
\end{align}
The expression for $s$ is given in series of small $d$.
\begin{align}
\label{x3x1plane_s}
s(d)&=\frac{\Gamma\left(\frac{1}{6}\right)}{2\sqrt{\pi}\ \Gamma\left(\frac{2}{3}\right)}\ d - \frac{5\Gamma\left(\frac{1}{3}\right)\Gamma\left(\frac{1}{6}\right)^6}{6912\sqrt{\pi^5}\ \Gamma\left(\frac{2}{3}\right)^6\Gamma\left(\frac{5}{6}\right)}\left(1+\frac{32}{3}\alpha^2+\frac{281}{1260}\varepsilon^2\alpha^2\right)d^5 \\ \nonumber
&+ \frac{\Gamma\left(\frac{1}{6}\right)^7}{448\sqrt{\pi^7}\Gamma\left(\frac{2}{3}\right)^7}\left(\frac{32}{3}\alpha^2+\frac{467}{420}\varepsilon^2\alpha^2\right)d^7+O(d^8)
\end{align}
$A_2$ in terms of $s$ upto 6th order is given by
\begin{align}
\label{x3x1plane_area}
\frac{A_2}{L_3L_1}&=2\int^\infty_{r_*} dr\ \frac{r^2}{f(r)\sqrt{N(r)}}\left(1-\frac{r^6_*}{r^6}\right)^{-\frac{1}{2}}\\ \nonumber
&=\int^\infty_1du\ \frac{2u^5}{\sqrt{u^6-1}}\frac{1}{s^3}\frac{1}{f(u/s)\sqrt{N(u/s)}}\\ \nonumber
&= 2I_4\frac{1}{s^2}+I_0\left(1+\frac{32}{3}\alpha^2+\frac{281}{1260}\varepsilon^2\alpha^2\right)s^2 -I_{-2}\left(\frac{32}{3}\alpha^2+\frac{467}{420}\varepsilon^2\alpha^2\right)s^4\\ \nonumber
&+I_{-4}\left(\frac{3}{4}+16\alpha^2+\frac{747}{280}\varepsilon^2\alpha^2\right)s^6+O(s^7)\\ \nonumber
&=\lim_{\delta\rightarrow 0}\frac{1}{\delta^2}-\frac{\sqrt{\pi}\Gamma\left(\frac{2}{3}\right)}{\Gamma\left(\frac{1}{6}\right)}\frac{1}{s^2}+\frac{\sqrt{\pi}\Gamma\left(\frac{1}{3}\right)}{6\ \Gamma\left(\frac{5}{6}\right)}\left(1+\frac{32}{3}\alpha^2+\frac{281}{1260}\varepsilon^2\alpha^2\right)s^2\\ \nonumber
&-\frac{\sqrt{\pi}\Gamma\left(\frac{2}{3}\right)}{\ \Gamma\left(\frac{1}{6}\right)}\left(\frac{32}{3}\alpha^2+\frac{467}{420}\varepsilon^2\alpha^2\right)s^4+\left(\frac{1}{4}+\frac{16}{3}\alpha^2+\frac{249}{280}\varepsilon^2\alpha^2\right)s^6 + O(s^7)
\end{align}
Finally, we get $A_2(d)$ by using the relations (\ref{x3x1plane_s}) and (\ref{x3x1plane_area}), which is 
\begin{align}
\frac{A_2}{L_3L_1}&= \lim_{\delta\rightarrow 0}\frac{1}{\delta^2}+C^{-2}_{31}\frac{1}{d^2}+C^2_{31}d^2+C^4_{31}d^4+C^6_{31}d^6+O(d^7)
\end{align}
where
\begin{align}
\nonumber
C^{-2}_{31}&=-\frac{32\sqrt{\pi^9}}{3\sqrt{3}}\frac{1}{\Gamma\left(\frac{1}{3}\right)^3\Gamma\left(\frac{1}{6}\right)^3}\\ \nonumber
C^2_{31}&=\frac{3}{320\sqrt{\pi^7}}\Gamma\left(\frac{1}{3}\right)^3\Gamma\left(\frac{1}{6}\right)^3\left(1+\frac{32}{3}\alpha^2+\frac{281}{1260}\varepsilon^2\alpha^2\right)\\ \nonumber
C^4_{31}&=-\frac{3\sqrt{3}}{28\sqrt{\pi^9}}\Gamma\left(\frac{1}{3}\right)^3\Gamma\left(\frac{1}{6}\right)^3\left(\alpha^2+\frac{467}{4480}\varepsilon^2\alpha^2\right)\\ \nonumber
C^6_{31}&=\frac{27}{16384\pi^9}\Gamma\left(\frac{1}{3}\right)^6\Gamma\left(\frac{1}{6}\right)^6\left(1-\frac{13\sqrt{3}}{1800\pi^{5/2}}\Gamma\left(\frac{1}{3}\right)^3\Gamma\left(\frac{1}{6}\right)^3\right)\\ \nonumber
&+\frac{9}{256\pi^9}\Gamma\left(\frac{1}{3}\right)^6\Gamma\left(\frac{1}{6}\right)^6\left(1-\frac{13\sqrt{3}}{1800\pi^{5/2}}\Gamma\left(\frac{1}{3}\right)^3\Gamma\left(\frac{1}{6}\right)^3\right)\alpha^2\\ \nonumber
&-\frac{6723}{1146880\pi^9}\Gamma\left(\frac{1}{3}\right)^6\Gamma\left(\frac{1}{6}\right)^6\left(1+\frac{3653\sqrt{3}}{4033800\pi^{5/2}}\Gamma\left(\frac{1}{3}\right)^3\Gamma\left(\frac{1}{6}\right)^3\right)\varepsilon^2\alpha^2,
\end{align}

\section{Equations and solutions for the Long cylinder}
\subsection{Solutions for $z_0(\rho)$, $z^{(\alpha)}(\rho)$ and $z^{(\varepsilon)}(\rho)$}
\label{appendixlabel-B1}
As we discussed in Sec.\ref{Holographic computation of entanglement entropy of a long cylinder with its radius}, we solve equation(\ref{equation-x-cylinder1}) by employing small $\alpha$ and $\varepsilon$ expansion upto leading order in $\alpha^2$ and $\varepsilon^2\alpha^2$. The form of the solution is given by
\begin{equation}
\label{zalphaepsilon}
z(\rho)=z_0(\rho)+\alpha^2 z^{(\alpha)}(\rho)+\varepsilon^2\alpha^2 z^{(\varepsilon)}(\rho),
\end{equation}
and each $z_0$,$z^{(\alpha)}$,$z^{(\varepsilon)}$ can be solved in the form of series solution in $\rho$. With such a form of the solution(\ref{zalphaepsilon}), we consider small $\alpha^2$ and $\varepsilon^2$ expansion of equation(\ref{equation-x-cylinder1}) and solve equations of zeroth order in $\alpha$ and $\varepsilon$, of leading order in $\alpha^2$, and $\varepsilon^2\alpha^2$.

{First,} let us examine the equation for $z_0$, which is zeroth order in $\alpha$ and $\varepsilon$, being given by
\begin{equation}
\sqrt{1+\frac{(z'_0)^2}{1-z_0^4}}-z_0^3\ \frac{\partial}{\partial \rho}\left(\frac{\rho}{z_0^3\sqrt{1+\frac{(z'_0)^2}{1-z_0^4}}}\right)=0,
\end{equation}
where prime denotes derivative with respect to $\rho$.
The solution for $z_0$ is {given by}

\begin{align}
\label{cylinder-zzero}
z_0&=z_*+\frac{3\left(z_*^4-1\right)}{4 z_*}\rho ^2+\frac{9\left(7 z_*^8-2 z_*^4-5\right)}{128 z_*^3}\rho ^4+\frac{3\left(10 z_*^{12}-13 z_*^8+16 z_*^4-13\right)}{128 z_*^5}\rho ^6\\ \nonumber
&+\frac{9\left(1275 z_*^{16}-2456 z_*^{12}-1354 z_*^8+7216 z_*^4-4681\right)}{131072
   z_*^7}\rho ^8\\ \nonumber
&+\frac{27\left(23275 z_*^{20}-49367 z_*^{16}+36878 z_*^{12}-184910 z_*^8+356407 z_*^4-182283\right)}{13107200 z_*^9}\rho ^{10}...{\rm upto\ O(\rho^{38})},
\end{align}
where we get the series solution upto order of $\rho^{38}$ but we just write the solution upto $O(\rho^{10})$.
{
%The prime in the equation (63) is derivative with respect to $\rho$.\\
%In fact, we solve the equation order by order in small $\rho$ up to $O(\rho^{38})$. 
To get the relation between $a$ and $z_*$, we apply the boundary condition(\ref{cylinder-z-solution-bc2}). The relation is obtained numerically.
%we addressed above and get the ratios of $a$ to $z_*$, which is given in Table 1 and 5.
Some results are given in Table \ref{Tabble-1}. In fact, to draw the graphs, we obtain more data points.
}
%%%%%%%%%%%%%%%%%%%%%%%%%%%%%%%%%%%%%%%%%%%%%
%%                Table 1                  %%
%%%%%%%%%%%%%%%%%%%%%%%%%%%%%%%%%%%%%%%%%%%%%
%\begin{table}[]
%\caption{\label{table-total-fit}In this table, we address the values of $a/z_*$, $z^{(\alpha)}_*$, and $z^{(\varepsilon)}_*$ obtained by best fitting of the graphs... }
%\end{table}

The equation of $z^{(\alpha)}(\rho)$ is given by
{
\begin{align}
\label{alpha-euqunavshf}
&\frac{1}{2}\frac{z_0'}{z_0^3}\frac{z_0'N_\alpha+2z^{(\alpha)\prime}+\frac{4z^{(\alpha)} z_0^3z_0'}{1-z_0^4}}{1-z_0^4+(z'_0)^2}\sqrt{1+\frac{(z'_0)^2}{1-z_0^4}}-3\frac{z^{(\alpha)}}{z_0^4}\sqrt{1+\frac{(z'_0)^2}{1-z_0^4}} \\ \nonumber
&= \frac{\partial}{\partial\rho}\left[-\frac{1}{2}\rho\ \frac{z_0'}{z_0^3}\frac{z_0'N_\alpha+2z^{(\alpha)\prime}+\frac{4z^{(\alpha)} z_0^3z_0'}{1-z_0^4}}{1-z_0^4+(z'_0)^2}\frac{1}{\sqrt{1+\frac{(z'_0)^2}{1-z_0^4}}}-3\rho\ \frac{z^{(\alpha)}}{z_0^4}\frac{1}{\sqrt{1+\frac{(z'_0)^2}{1-z_0^4}}}\right],
\end{align}
}
where
\begin{align}
N_\alpha=\frac{32}{3}\frac{z_0^4}{1+z_0^2}.
\end{align}
We note that the equation is that of leading order in $\alpha^2$.
We put the zeroth order solution, $z_0$ that we obtained previously into the equation(\ref{alpha-euqunavshf}) and get the solution of $z^{(\alpha)}(\rho)$ which is also a series solution in small $\rho$. The solution, $z^{(\alpha)}(\rho)$ is given by
{
\begin{align}
\label{cylinder-zalpha}
z^{(\alpha)} &= z_*^{\alpha}
+ \frac{9 z_*^4 z_*^{\alpha }+3 z_*^{\alpha }-32 z_*^7+32 z_*^5}{4 z_*^2} \rho ^2
\\ \nonumber
&+ \frac{315 z_*^8 z_*^{\alpha }-18 z_*^4 z_*^{\alpha }+135 z_*^{\alpha }-1920 z_*^{11}+1344 z_*^9+768 z_*^7-192 z_*^5}{128 z_*^4} \rho ^4
\\ \nonumber
&+ \frac{1}{128 z_*^6} \left( 210 z_*^{12} z_*^{\alpha }-117 z_*^8 z_*^{\alpha }-48 z_*^4 z_*^{\alpha }+195 z_*^{\alpha } \right. \\ \nonumber
&\left.-2000 z_*^{15}+960 z_*^{13}+2272 z_*^{11}-832 z_*^9-912 z_*^7+512 z_*^5\right) \rho ^6...+{\rm \ upto \ O(\rho^{30})}
\end{align}
}

We also get the equation and the solution for $z^{(\varepsilon)}$, being given by
{
\begin{align}
&\frac{1}{2}\frac{z_0'}{z_0^3}\frac{z_0'(N_\varepsilon+2F)+2z^{(\varepsilon)\prime}+\frac{4z^{(\varepsilon)} z_0^3z_0'}{1-z_0^4}}{1-z_0^4+(z'_0)^2}\sqrt{1+\frac{(z'_0)^2}{1-z_0^4}}-3\frac{z^{(\varepsilon)}}{z_0^4}\sqrt{1+\frac{(z'_0)^2}{1-z_0^4}} \\ \nonumber
&= \frac{\partial}{\partial\rho}\left[-\frac{1}{2}\rho\ \frac{z_0'}{z_0^3}\frac{z_0'(N_\varepsilon+2F)+2z^{(\varepsilon)\prime}+\frac{4z^{(\varepsilon)} z_0^3z_0'}{1-z_0^4}}{1-z_0^4+(z'_0)^2}\frac{1}{\sqrt{1+\frac{(z'_0)^2}{1-z_0^4}}}-3\rho\ \frac{z^{(\varepsilon)}}{z_0^4}\frac{1}{\sqrt{1+\frac{(z'_0)^2}{1-z_0^4}}}\right],
\end{align}
}
where
\begin{align}
N_\varepsilon&=\frac{4}{9}\ z_0^4\left(\frac{281}{560}-\frac{2z_0^6+6z_0^4+3z_0^2}{2\ (z_0^2+1)^4}\right),\\
F&=\frac{1}{18}\ z_0^6\frac{z_0^2-2}{(z_0^2+1)^4}.
\end{align}
%(\ref{cylinder-zzero}), (\ref{cylinder-zalpha}) and (\ref{cylinder-zepsilon})
{The solution of $z^{(\varepsilon)}$ is given by}
{
\begin{align}
\label{cylinder-zepsilon}
z^{(\varepsilon)} &= z_*^{\varepsilon }
+ \frac{1}{1680 z_*^2 \left(z_*^2+1\right){}^3} \left(3780 z_*^{10} z_*^{\varepsilon }+11340 z_*^8 z_*^{\varepsilon }+12600 z_*^6 z_*^{\varepsilon }+7560 z_*^4 z_*^{\varepsilon}+3780 z_*^2 z_*^{\varepsilon}\right. \\ \nonumber
&\left.+1260 z_*^{\varepsilon}-281z_*^{15}-283 z_*^{13}+418 z_*^{11}+142 z_*^9-277 z_*^7+281 z_*^5 \right)\rho ^2
\\ \nonumber
&+ \frac{1}{8960 z_*^4 \left(z_*^2+1\right){}^3} \left(22050 z_*^{14} z_*^{\varepsilon }+66150 z_*^{12} z_*^{\varepsilon }+64890 z_*^{10} z_*^{\varepsilon }+18270 z_*^8 z_*^{\varepsilon }+5670 z_*^6 z_*^{\varepsilon}\right. \\ \nonumber
&\left.+27090 z_*^4 z_*^{\varepsilon }+28350z_*^2 z_*^{\varepsilon }+9450 z_*^{\varepsilon }-3653 z_*^{19}-5359 z_*^{17}+4875 z_*^{15}+8009 z_*^{13}-3179 z_*^{11}\right. \\ \nonumber
&\left.-4049 z_*^9+3637 z_*^7-281 z_*^5\right)\rho ^4
\\ \nonumber
&+ \frac{1}{53760 z_*^6 \left(z_*^2+1\right){}^3} \left(88200 z_*^{18} z_*^{\varepsilon }+264600 z_*^{16} z_*^{\varepsilon }+215460 z_*^{14} z_*^{\varepsilon }-59220 z_*^{12} z_*^{\varepsilon }-167580 z_*^{10} z_*^{\varepsilon}\right. \\ \nonumber
&\left.-109620 z_*^8 z_*^{\varepsilon}+21420 z_*^6 z_*^{\varepsilon }+225540 z_*^4 z_*^{\varepsilon }+245700 z_*^2 z_*^{\varepsilon }+81900 z_*^{\varepsilon }-31191 z_*^{23}-58573 z_*^{21}\right.
\\ \nonumber
&\left.+40445 z_*^{19}+121103 z_*^{17}+3111z_*^{15}-89251 z_*^{13}-33693 z_*^{11}+61985 z_*^9-18432 z_*^7+4496 z_*^5\right) \rho ^6\\ \nonumber
&+{\rm \ upto \ O(\rho^{30})}
\end{align}
}

\subsection{Black brane solution and entanglement temperature}
\label{appendixlabel-B2}
In this subsection, we find the solutions of $z_0^A$ and $z_0^B$. In this case, $f(r)$ and $N(r)$ are given by
\begin{equation}
\nonumber
f=1,\ N(r,\xi)=r^2-\frac{\xi}{r^2},
\end{equation}
and $z_0$ will have a form of
\begin{equation}
\nonumber
z_0=z_0^A+\xi z_0^B+O(\xi^2),
\end{equation}
where $\xi$ is a bookkeeping parameter and later we take $\xi=1$.
The equation of $z_0$ is given by
\begin{equation}
\sqrt{1+\frac{(z'_0)^2}{1-\xi z_0^4}}-z_0^3\ \frac{\partial}{\partial \rho}\left(\frac{\rho}{z_0^3\sqrt{1+\frac{(z'_0)^2}{1-\xi z_0^4}}}\right)=0.
\end{equation}
The solutions for $z_0^A$ and $z_0^B$ are obtained by employing power expansion order by order in $\xi$ upto its first order. We get series solutions for $z_0^A$ and $z_0^B$ in $\rho$ upto $O(\rho^{75})$, which are given by
\begin{align}
z_0^A&= z_*^A -\frac{3}{4 (z_*^A)}\rho^2-\frac{45}{128(z_*^A)^3}\rho^4-\frac{39}{128 (z_*^A)^5}\rho ^6+\cdots,
\\ \nonumber
z_0^B&= z_*^B +\frac{3 (z_*^A)^5+3 z_*^B}{4 (z_*^A)^2}\rho^2 +\frac{-18 (z_*^A)^5+135 z_*^B}{128 (z_*^A)^4}\rho^4+\frac{48 (z_*^A)^5+195 z_*^B}{128 (z_*^A)^6}\rho^6+\cdots.
\end{align}
The boundary condition $z_0^A (a) = 0$ gives a constant ratio of $a$ to $z_*^A$ as
\begin{equation}
\frac{a}{z_*^A} = 0.789541.
\end{equation}
With the given values of $a$ from the previous computation, the boundary condition $z_0^B (a) = 0$ gives the value of $z_*^B$. With these values we obtain
\begin{equation}
\lim_{a\rightarrow 0}\frac{\gamma^{(\xi)}}{a^3}= 0.73228506.
\end{equation}
The solution of $\gamma^{(\xi)}(a)$ is given explicitly in the next subsection, upto 6th order.

\subsection{Computation of the surface area for cylinder}
\label{appendixlabel-B3}
Defining $r\equiv z^{-1}(\rho)$, the integral for the surface area of the cylinder becomes
\begin{equation}
\label{cylinder-area-integral-in-z}
A^{cy}_1 = 2 \pi L_1 \int^0_a z^{-3}(\rho) \rho \left(\frac{\partial z}{\partial \rho}\right) d\rho \sqrt{\frac{1}{\left(\frac{\partial z}{\partial \rho}\right)^2}+ \frac{1}{z^2N(z^{-1})f^2(z^{-1})}}.
\end{equation}
%which yields the equation of motion
%\begin{equation}
%\label{cylinder-z-eom}
%\sqrt{1+\frac{(\frac{dz(\rho)}{d\rho})^2}{z^2f(z^{-1})N(z^{-1})}}-z^3(\rho)\ \frac{\partial}{\partial \rho}\left(\frac{\rho}{z^3(\rho)\sqrt{1+\frac{(\frac{dz(\rho)}{d\rho})^2}{z^2f(z^{-1})N(z^{-1})}}}\right)=0.
%\end{equation}
%Now, we are ready to compute the minimal area of the cylinder. Since the area is divergent as we discussed, we concentrated on the finite part by removing the UV-divergences near AdS boundary. Then, the finite part of the area is given by
%\begin{align}
%\mathcal A^{cy\ (finite)}_x &= 2\pi L_x \left[\int^{z_*}_\delta dz\ z^{-3} \rho (z) \sqrt{\frac{1}{z^2f^2(z^{-1})N(z^{-1})}+\left(\frac{\partial\rho}{\partial z}\right)^2} - \int_\delta dz \left(\frac{a}{z^3}-\frac{1}{8az}\right)\right]
%\\ \nonumber
%&= 2\pi L_x \left[ \int^{z_*}_\delta dz  z^{-3} \left\{\rho (z) \sqrt{\frac{1}{z^2f^2(z^{-1})N(z^{-1})}+\left(\frac{\partial\rho}{\partial z}\right)^2} -a+\frac{z^2}{8a}\right\} \right.
%\\ \nonumber
%&\left. - \frac{a}{2z^2_*}-\frac{1}{8a}\log z_* \right]
%\end{align}
We expand the integrand upto leading order of $\xi$, $\alpha^2$, and $\varepsilon^2 \alpha^2$, as we similarly did for $z(\rho)$:
\begin{equation}
\nonumber
A_1^{cy} = 2\pi L_1 \left( \gamma^{(AdS)} + \xi\gamma^{(\xi)} + \alpha^2\gamma^{(\alpha)} + \varepsilon^2\alpha^2\gamma^{(\varepsilon)} \right).
\end{equation}
The solutions of $z^{(\alpha)}(\rho)$, $z^{(\varepsilon)}(\rho)$, $z_0^A(\rho)$ and $z_0^B(\rho)$ are given in Appendix \ref{appendixlabel-B1} and \ref{appendixlabel-B2}. With these solutions we may expand the integrands in terms of $\rho$ and integrate them to get $\gamma^{(\xi)}$, $\gamma^{(\alpha)}$, and $\gamma^{(\varepsilon)}$:
\begin{align}
\gamma ^{(\xi)} &=
- \frac{3 z_*^B}{2 (z_*^A)^4} a^2
- \frac{27 \{ (z_*^A)^5 + 5 z_*^B \} }{32 (z_*^A)^6} a^4
- \frac{45 \{ 3 (z_*^A)^5 + 14 z_*^B \} }{64 (z_*^A)^8} a^6
+ \rm{upto}\ O(a^{75}),
\\
\gamma ^{(\alpha)} &= - \frac{3 z_* ^{(\alpha)}}{2z_* ^4} a^2
+ \frac{9(-32z_*^5 + 32z_*^7 -15z_* ^{(\alpha)} + 3z_*^4 z_* ^{(\alpha)} )}{32z_*^6} a^4
\\ \nonumber
&- \frac{15( 96z_*^5 - 80 z_*^7 - 64 z_*^9 + 48 z_*^11 + 42z_* ^{(\alpha)} - 27 z_*^4 z_* ^{(\alpha)} - 3z_*^8 z_* ^{(\alpha)} )}{64 z_*^8} a^6
+ \rm{upto}\ O(a^{30}),
\\
\gamma ^{(\varepsilon)} &=
- \frac{3 z_* ^{(\varepsilon)} }{2 z_*^4} a^2
\\ \nonumber
&+\frac{3}{4480 z_*^6 (z_*^2+1)^3} \left\{ -281 z_*^5 + 277 z_*^7 -142 z_*^9 -418 z_*^{11} + 283 z_*^{13} + 281 z_*^{15} \right.
\\ \nonumber
&\qquad\qquad\qquad\qquad
+ z_* ^{(\varepsilon)} \left. ( - 6300 - 18900 z_*^2 -17640 z_*^4 -2520 z_*^6  +3780 z_*^8  + 1260 z_*^{10} ) \right\}\ a^4
\\ \nonumber
&+ \frac{1}{1792 z_*^8 (1+z_*^2)^3} \left\{ -843z_*^5 + 271z_*^7 + 1115 z_*^9 -1671 z_*^{11} + 11 z_*^{13} + 1401 z_*^{15} - 3 z_*^{17} - 281 z_*^{19} \right.
\\ \nonumber
&\qquad\qquad\qquad\qquad
+  z_* ^{(\varepsilon)} ( -17640 + 52920 z_*^2 -41580 z_*^4 + 16380 z_*^6 +35280 z_*^8 + 15120 z_*^{10} + 3780 z_*^{12}
\\ \nonumber
&\qquad\qquad\qquad\qquad\qquad\quad
\left. + 1260 z_*^{14}  ) \right\} a^6
\\ \nonumber
&+ \rm{upto}\ O(a^{30}).
\end{align}
\end{appendices}

\end{document}